\documentclass[twocolumn]{aastex631}  %twocolumn
\usepackage{enumitem}
\usepackage{xcolor}
\usepackage{tablefootnote}
\usepackage{bm}
\usepackage{lineno}
\hypersetup{backref=true,pagebackref=true,  hyperindex=true,colorlinks=blue,breaklinks=true,  urlcolor=violet,linkcolor= red, bookmarks=true, 	bookmarksopen=true,  filecolor=cyan, citecolor=blue, linkbordercolor=blue}
\shorttitle{Self-Lensing/Eclipsing Signals by the TESS telescope}
\shortauthors{Sajadian and Afshordi}

\begin{document}
%\linenumbers
\title{Simulating Self-Lensing and Eclipsing Signals due to Detached Compact Objects in the TESS Light Curves}

\author[0000-0002-0167-3595]{Sedighe Sajadian}
\affiliation{Department of Physics, Isfahan University of Technology, Isfahan 84156-83111, Iran, \url{s.sajadian@iut.ac.ir}}

\author[0000-0002-9940-7040]{Niayesh Afshordi}
\affiliation{Waterloo Centre for Astrophysics, University of Waterloo, Waterloo, Ontario N2L 3G1, Canada}
\affiliation{Department of Physics and Astronomy, University of Waterloo, Waterloo, Ontario N2L 3G1, Canada}
\affiliation{Perimeter Institute for Theoretical Physics, Waterloo, Ontario N2L 2Y5, Canada}

\begin{abstract}
A fraction of Galactic stars have compact companions which could be white dwarfs (WDs), neutron stars (NSs) or stellar-mass black holes (SBHs). In a detached and edge-on binary system including a main-sequence star and a compact object (denoted by WDMS, NSMS, and BHMS systems), the stellar brightness can change periodically due to self-lensing or eclipsing features. The shape of a self-lensing signals is a degenerate function of stellar radius and compact object's mass because the self-lensing peak strongly depends on the projected source radius normalized to Einstein radius. Increasing the inclination angle $i$ changes the self-lensing shape from a strict top-hat model to one with slow-increasing edges. We simulate stellar light curves due to these binary systems which are observed by NASA's Transiting Exoplanet Survey Satellite (TESS) telescope and evaluate the efficiencies to detect their periodic signatures using two sets of criteria (i)SNR$>3$ and $N_{\rm{tran}}>1$ (Low-Confidence, LC), and (ii) SNR$>5$ and $N_{\rm{tran}}>2$ (High-Confidence, HC). The HC efficiencies for detecting WDMS, NSMS, and BHMS systems with the inclination angle $i<20^{\circ}$ during different time spans are $5$-$7\%$, $4.5$-$6\%$, and $4$-$5\%$, respectively. Detecting lensing-induced features is possible in only $\lesssim3\%$ and $\lesssim33\%$ of detectable WDMS and NSMS events. The detection efficiencies for closer source stars with higher priorities are higher and drop to zero for $b\gtrsim R_{\star}$, where $b\simeq \tan(i) a$ is the impact parameter($a$ is the semi-major axis). We predict the numbers of WDs, NSs, and SBHs that are discovered from the TESS Candidate Target List stars are $15$-$18$, $6$-$7$, and $<1$.
\end{abstract}


%%%%%%%%%%%%%%%%%%%%%%%%%%%%%%%%%
\section{Introduction}
The NASA's Transiting Exoplanet Survey Satellite (TESS \footnote{\url{https://tess.mit.edu/}}, \citet{2014SPIEricker}) telescope observed (and observes) stars in the Candidate Target List (CTL, \citet{2018AJStassun,2019AJStassun}) with a $2$-min cadence and an accuracy better than $60$ ppm on hourly timescales. Although its main goal from these observations is to discover Earth-size planets transiting bright stars in the solar neighborhood, its observing strategy is uniquely matched to capture other types of periodic and weak variations in stellar light curves. 

\noindent For instance, stellar light curves from edge-on and detached binary systems including main-sequence stars and compact objects, i.e., white dwarfs (WDs), neutron stars (NSs), and stellar-mass black holes (SBHs), have potentially self-lensing, eclipsing (blocking the compact objects' brightness by their stellar companions), and occultation (blocking the light of stellar images by their compact companions) signals which all of them are periodic. Self-lensing refers to the lensing effect on the flux of the luminous object by its compact companion in edge-on systems \citep[see, e.g.,][]{1995ApJGould}. In this regard, simulating and numerically studying self-lensing/eclipsing/occultation features that are realizable in the TESS data have several advantages as (i) revealing the importance of searching these signals in the TESS data, (ii) learning the machines to extract the real events from a huge ensemble of the TESS data archive which is growing up with time, and (iii) evaluating the known models describing binary systems, NSs, and SBHs through comparing their results with ones from the real observations.  

\noindent In this work, we therefore simulate these periodic features and study their properties and detectability in the TESS data. However, \citet{2019ApJMasuda} roughly evaluated the number of detectable BHs in the mass range $[1,~100] M_{\odot}$ in tight and detached orbits around stars with the orbital period $[0.3,~30]\rm{days}$ through either self-lensing signals or phase-curve variations of stellar light curves. They injected these photometric signals in spotted stellar light curves detected by the Kepler telescope and concluded that $10$-$100$ BHs can be detected by searching $10^{5}-10^{6}$ stellar light curves.

\noindent Also, \citet{2021Wiktorowicz} studied detection of BHs/NSs through their self-lensing signals in a synthetic ensemble of binary systems and reported a higher number for detectable BHs. Here, we extend their works by (i) considering binary systems including different types of compact objects (i.e., WDs, NSs, and BHs) separately, (ii) modeling eclipsing and occultation in addition to self-lensing signals, (iii) generating synthetic light curves and data points based on the real observations by the TESS telescope, and (iv) taking their companion stars from the TESS CTL targets.

Here, we first review the known properties of binary systems including main-sequence stars and compact objects in three following paragraphs.  

{\bf WD main-sequence (WDMS) binaries:}~ It is predicted that the number of WDMS binaries in our galaxy is $10^{7}$-$10^{8}$ which depends on their mass-ratio distribution. The number of detached WDMS binaries is more than that of interacting ones by more than one order of magnitude \citep[e.g., see][]{2004AAwillems}, nevertheless detecting interacting WDMS binaries is easier and plausible via either spectroscopic observations or variability surveys. For instance, the Sloan Digital Sky Survey (SDSS, \citet{2000SDSSyork}) telescope discovered more than $3200$ WDMS binaries \citep[see, e.g., ][]{2009AAHeller,2016MNRASRebassa}. Additionally, $\sim320,000$ (either candidate or confirmed) WDs in binary systems have been reported in the Galaxy Evolution Explorer (GALEX, \citet{2005ApJGALex}) and the Gaia Data Release 3 (GDR3, \citet{2023AAGaiaDR3}) \citep{2011MNRASGalexWD,2021MNRASWDGaia}. For detecting detached WDMS binaries, a common method is spectroscopic observations because they can make a combined spectrum due to one star and one WD. If these systems are edge-on as seen by the observer, additionally eclipsing, self-lensing signals or variation in stellar radial velocities are realizable through precise photometric and spectroscopic observations \citep[][]{2017MNRASKorol}. 

{\bf NS main-sequence (NSMS) binaries:}~Another type of compact object is an NS which could be a companion for a main-sequence star in a binary system. A NS is formed after the collapse of a massive star with an initial mass higher than $8~M_{\odot}$, whereas more massive stars with initial masses higher than $\sim 20~M_{\odot}$ are converted to SBHs after their gravitational collapse. Isolated NSs can be discovered through radio emissions, the so-called pulsars \citep[e.g., ][]{2022ApJRadioPulsar}. NSs and SBHs in interacting binary systems with main-sequence stars, the so-called $X$-ray binaries, are discernible through $X$-ray emissions owing to mass transferring from their stellar counterparts. Depending on the dominant mechanism for mass transferring toward the compact objects, these binaries are divided into two subclasses: (i) low-mass $X$-ray binaries (with Roche-Lobe overflow), and (ii) high-mass $X$-ray binaries (with wind-accreting) \citep[see, e.g.,  ][]{1974Xraybinary,2002ApJPfahl, 2017bookCasares}. Around $4\%$ of all discovered NSs are in binary systems \citep{2006bookXray}, and our galaxy hosts up to one billion NSs, whereas only $\sim 4,000$ NSs have been discovered up to now.

{\bf SBH main-sequence (BHMS) binaries:}~Black holes with masses $\lesssim 100~M_{\odot}$ are the so-called stellar-mass black holes. The lightest SBH was discovered up to now has the mass $\simeq 3$-$3.3M_{\odot}$ which was located in the mass-gap between NSs and SBHs \citep{2019SciThompson,2021MNRASJayasinghe,2010ApJOzel,2011ApJFarr}. Although the number of SBHs in our galaxy is predicted to be more than $10$ million, up to now only $72$ SBHs were confirmed mostly through $X$-ray transients from their accretion disks in interacting BHMS binaries \footnote{\url{https://www.astro.puc.cl/BlackCAT/transients.php}} \citep{2016AABlackCat}. $20$ of these discovered SBHs were confirmed dynamically by discerning periodic variations in the radial velocities of their luminous companions.

In addition to spectroscopic and variability surveys, and $X$-ray observations to capture compact objects in binary systems with main-sequence stars, there are other channels for discovering these objects, e.g., (i) based on their gravitational impacts which is the so-called self-lensing effect, (ii) eclipsing signals, (iii) ellipsoidal variation of stellar brightness, (iv) Doppler boosting, etc \citep[e.g., see,][]{2019ApJMasuda,2022ApJsorebella}. Two last effects happen for massive compact companions and small orbital radii. Unlike other types of lensing, including microlensing, strong lensing and weak lensing which all are not repeatable, a self-lensing signal is periodic and its period is exactly equal to the orbital period of the binary system. 

\noindent Up to now, five self-lensing/eclipsing binary systems containing white dwarfs and main-sequence stars were discovered through photometric observations by the Kepler telescope \citep{KruseAgol2014,2018AJKawahara,2019ApJLMasuda}. All of these systems are wide binaries with orbital periods from $88$ to $728$ days. The next generation of the Kepler telescope is the TESS telescope which was launched on $18$ April $2018$. During its two-year primary mission, it covered $85\%$ of the sky by dividing it into $26$ sectors. The area of each sector is $24\times 90$ square degrees, and sectors have some overlapped parts over the ecliptic poles. Each sector is observed during two $13.7$-day observing periods with a one-day gap in the middle. The TESS pixel scale is $21$ arc-second which leads to a high photometric accuracy better than $60$ ppm (parts per million) for brightest stars on hourly time scales. This telescope could also re-cover self-lensing signals in a binary system originally discovered by the Kepler telescope, i.e., KIC~12254688~\citep{2024ApJSorebella}.

In this work, we simulate possible self-lensing and eclipsing signals due to compact objects in stellar fluxes from detached binary systems as seen by the TESS telescope, to (i) study the characteristics of self-lensing signals and (ii) estimate the TESS efficiency for detecting them. We also investigate how this detection efficiency depends on the physical parameters of compact objects, source stars, and binary orbits. We finally estimate the numbers of WDs, NSs, and SBHs that are realizable through their self-lensing/eclipsing signals in the photometric data of the TESS CTL targets.

The outline of this paper is as follows. In Section \ref{sec1}, we explain the details of simulating self-lensing, occultation, and eclipsing signals and then discuss on their characteristics as a function of models' parameters. In Section \ref{sec2}, we explain the details of Monte Carlo simulations from self-lensing, occultation, and eclipsing light curves due to detached and edge-on binary systems while we assume that these events are observed by the TESS telescope. We extract the detectable events based on two sets of criteria. In Section \ref{sec3}, we explain the results and conclusions.

\section{Binary Systems Contain Compact Objects and Main-Sequence Stars}\label{sec1}
To simulate a stellar light curve from a binary system including a compact object and a main-sequence star, in the first step, we simulate a binary orbit as explained in Subsection \ref{sub1}. By assuming that its orbital plane is edge-on as seen by the observer, in the next step we calculate its self-lensing, occultation, and eclipsing signals. All details are explained in Subsection \ref{sub2}. We then discuss the characteristics of self-lensing signals as functions of orbital properties and parameters of compact objects in Subsection \ref{sub3}.  
 
\subsection{Simulating a Stellar Orbit Around a Compact Object}\label{sub1}
Let's consider a detached binary system containing two companions: a star and a compact object with masses $M_{\star}$, and $M_{\rm c}$ ($c$ can be either WD, or NS, or SBH). We assume this system is isolated and there is no external force. Hence, their center of mass (CM) moves with a constant velocity and both components rotate over elliptical orbits so that these orbits have a common barycenter on their CM's location. In the CM coordinate system, their CM is fixed and the star with respect to the compact object moves over an elliptical orbit. We assume that the semi-major axis of this elliptical orbit is $a$. The period of this elliptical orbit is given by the third Kepler's law as follows: 
\begin{eqnarray}
T=\frac{2 \pi}{\sqrt{G (M_{\star}+M_{\rm c})}}~a^{3/2}.
\end{eqnarray} 
This elliptical orbit is characterized as \citep[See, e.g., ][]{1998Dominik}:
\begin{eqnarray}
x&=&a~\big(\cos \xi- \epsilon\big), \nonumber \\
y&=&a~\sin \xi~\sqrt{1-\epsilon^{2}},
\end{eqnarray} 
where, $\epsilon$ is the orbital eccentricity. $\xi$, the so-called eccentric anomaly, is given by Kepler's equation, i.e., $\phi= \xi-\epsilon~\sin \xi$, where $\phi=\omega~(t-t_{\rm p})$ is called the mean anomaly, and $\omega=2 \pi\big/T$ is the angular velocity. Also, $t_{\rm p}$ is a characteristic time which indicates the time of crossing the orbital periapsis point. Hence, $(x,~y)$ defines the orbital plane of the star around the compact object. In the simulation, we solve Kepler's equation numerically using:  
\begin{eqnarray}
\xi=\phi+ \sum_{i=1}^{\infty}\frac{2}{\pi}~\sin (i \phi)~\mathcal{B}_{i}(i \epsilon),
\end{eqnarray}
where, $\mathcal{B}_{i}$ is the known Bessel function of $i$th order.

We define the observer's coordinate system, $(x_{\rm o},~y_{\rm o},~z_{\rm o})$, where $x_{\rm o}$ is toward the observer, and $(y_{\rm o},~z_{\rm o})$ defines the sky plane. We then convert the CM coordinate system to the observer coordinate system using two projection angles, i.e., $\theta$ and $i$. $\theta$ is the angle between the minor axis of this elliptical orbit and the sky plane, and $i$, the so-called inclination angle, is the angle between the line of sight toward the observer and the orbital plane. Hence, the orbital components in the observer coordinate system are given by: 
\begin{eqnarray}
x_{\rm o}&=&\cos i~\big(-y \sin \theta + x \cos \theta \big), \nonumber\\
y_{\rm o}&=&y \cos \theta + x \sin \theta, \nonumber\\
z_{\rm o}&=&-\sin i~\big(-y \sin \theta +x \cos \theta \big).
\end{eqnarray}
For stellar orbits with small inclination angles (the so-called edge-on ones) the stellar light is magnified by the compact object once an orbital period and whenever the star is passing behind it. We explain the details of calculating self-lensing, occultation and eclipsing signals in the following subsection. 

\begin{figure*}
\centering
\includegraphics[width=0.45\textwidth]{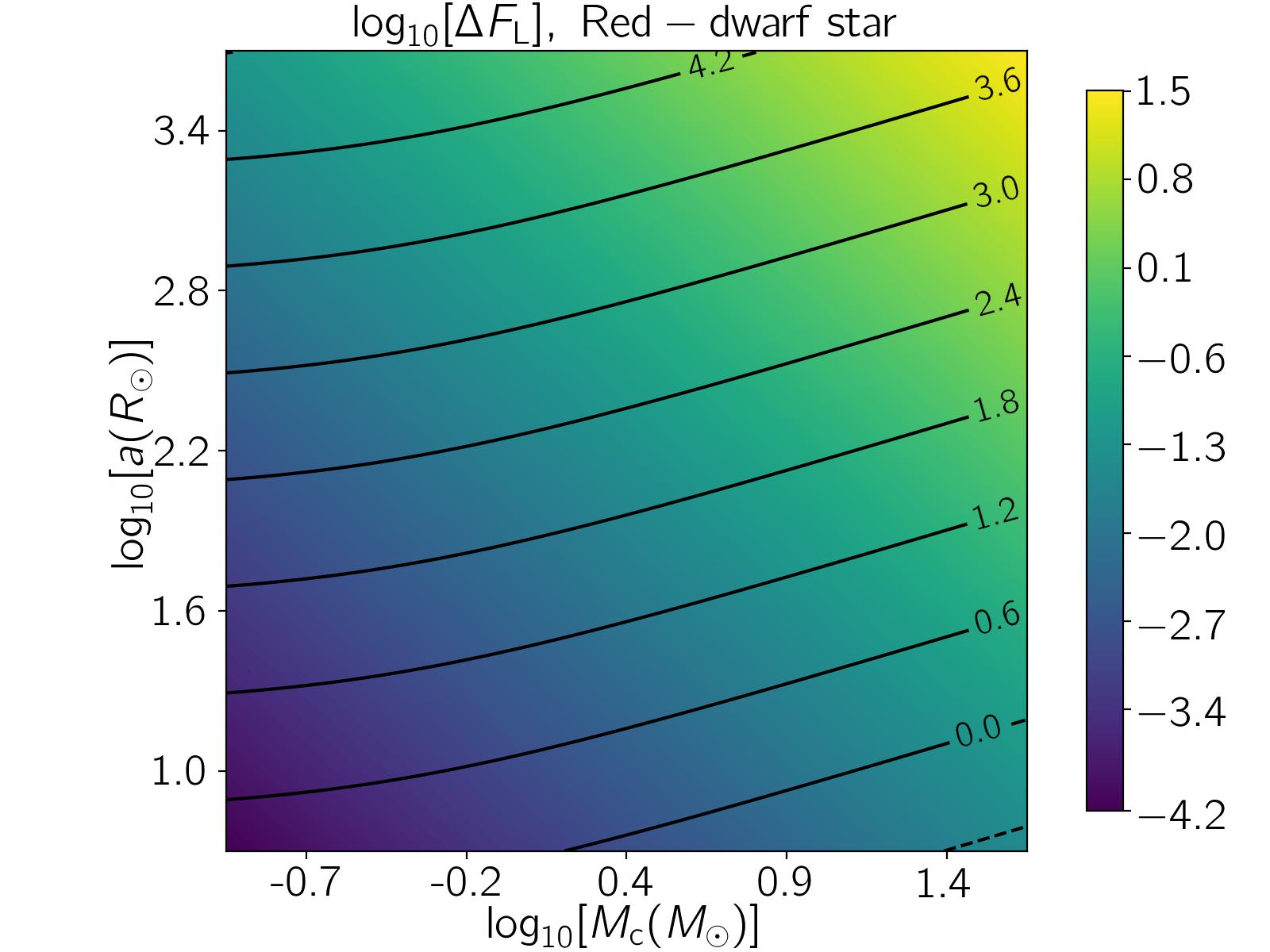}
\includegraphics[width=0.45\textwidth]{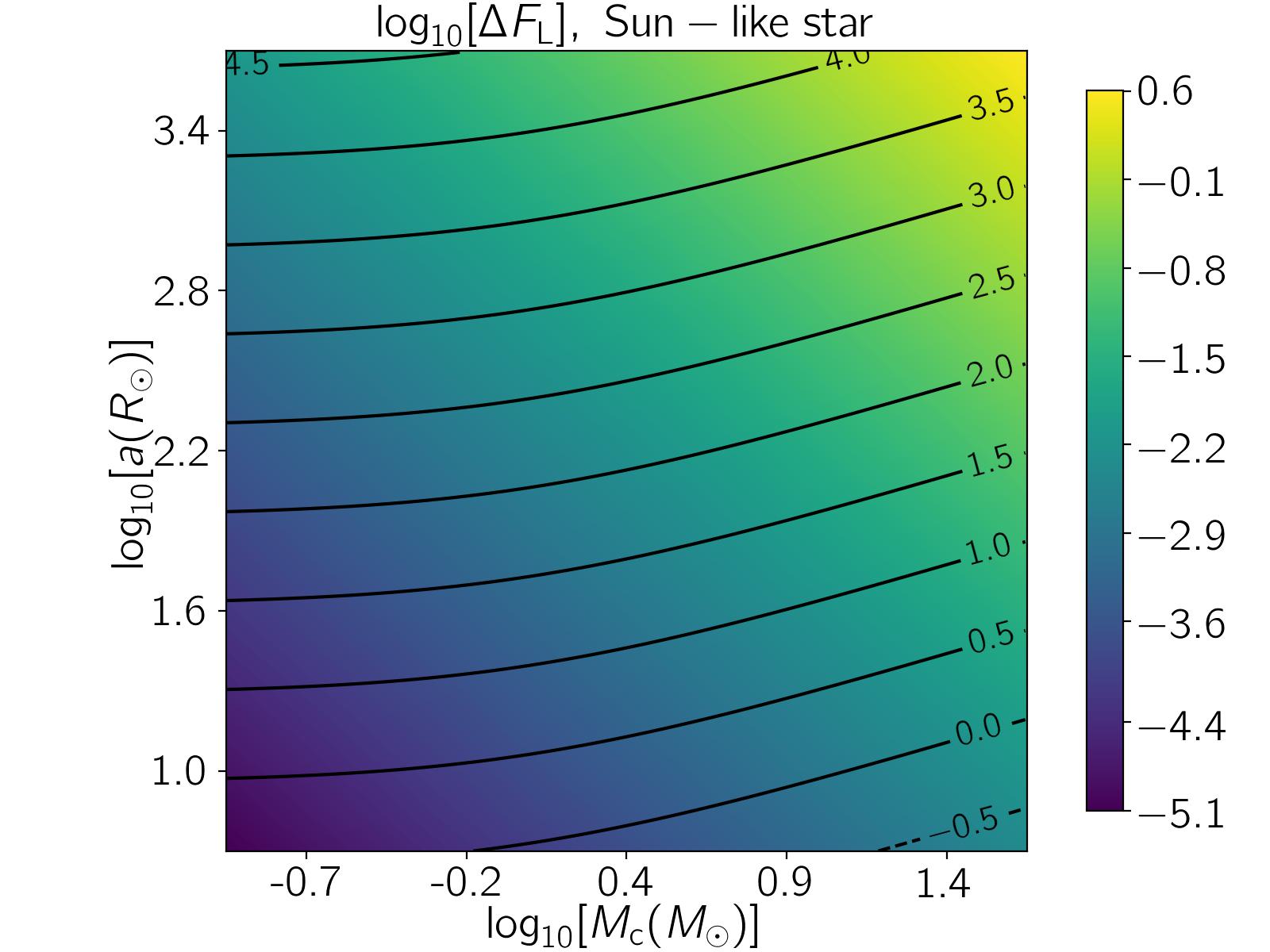}
\includegraphics[width=0.45\textwidth]{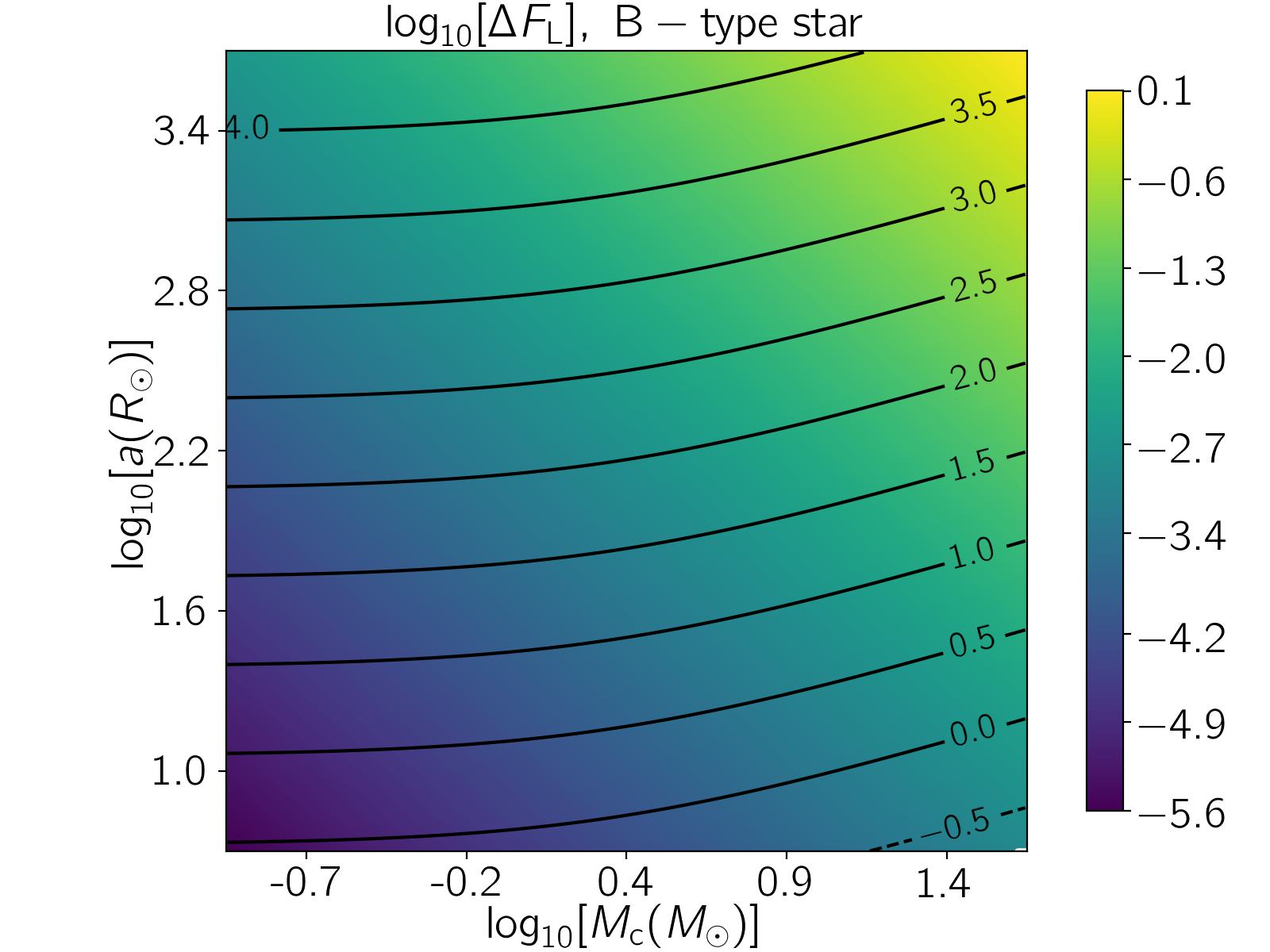}
\includegraphics[width=0.45\textwidth]{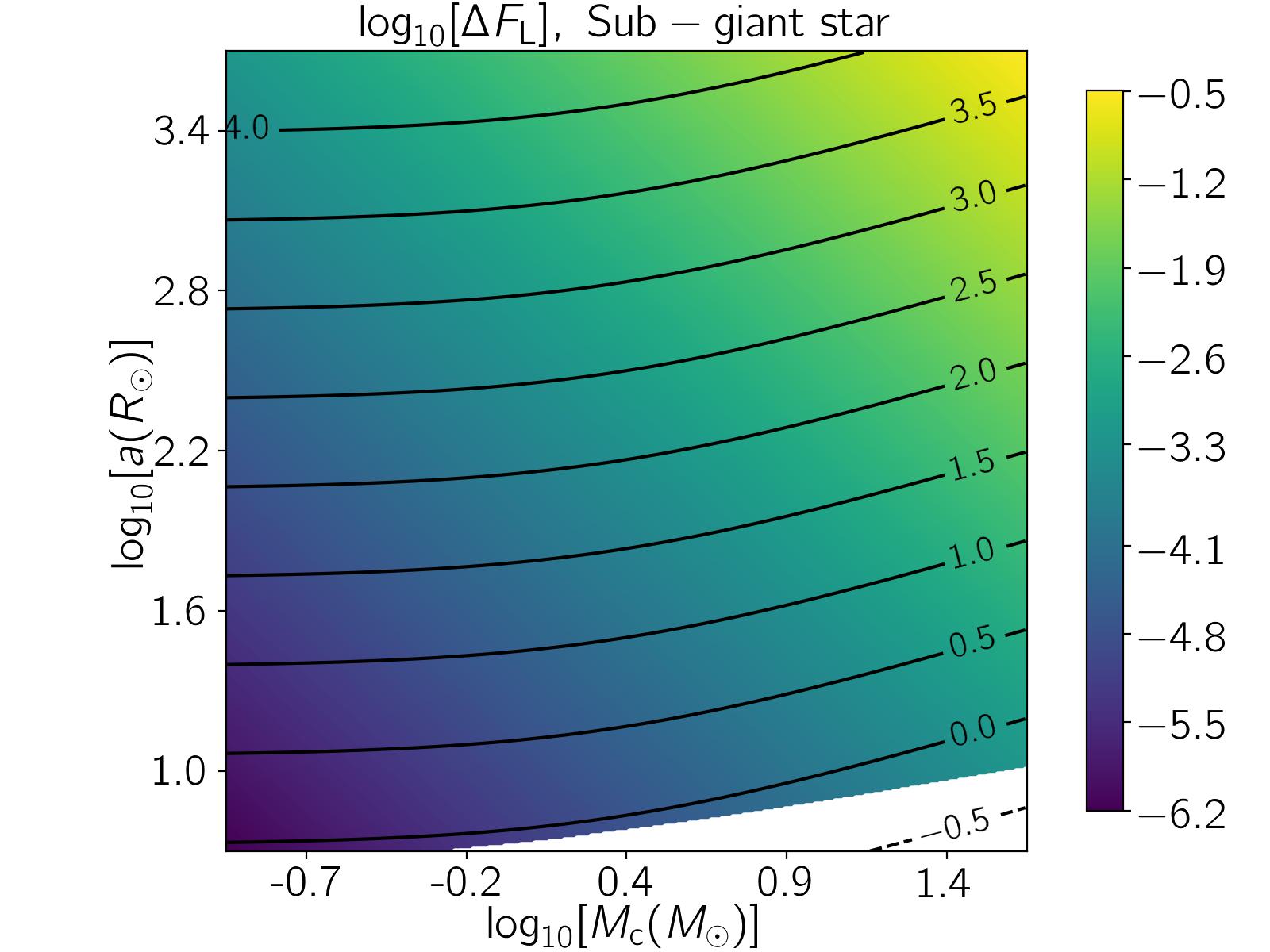}
\caption{Maps of $\log_{10}[\Delta F_{\rm L}]$ in detached and edge-on binary systems over the 2D space of $\log_{10}[M_{\rm c} (M_{\odot})]- \log_{10}[a (R_{\sun})]$. We consider four different source stars in binary systems include: (i) a red dwarf with $M_{\star}=0.3 M_{\odot}$ and $R_{\star}=0.35 R_{\odot}$, (ii) a Sun-like star, (iii) a B-type star with $M_{\star}=2 M_{\odot}$, and $R_{\star}=1.8 R_{\odot}$, and (iv) a sub-giant star with $M_{\star}=2 M_{\odot}$, and $R_{\star}=3.6 R_{\sun}$. The contours display $\log_{10}[T(\rm{days})]$. The interacting binaries with $D_{\rm{RL}}<R_{\star}$ (see, Eq. \ref{rochelobe}) are excluded and covered with white color on maps.} \label{maps}
\end{figure*}

\subsection{Simulating Self-Lensing, Occultation, and Eclipsing Signals} \label{sub2}
In the stellar orbit around the compact object (the lens object), there is a phase angle $\Phi$ which is the angle between two lines of sight from the source star: one toward the compact object and another one toward the observer. It is calculated by $\cos \Phi=-x_{\rm o}\big/D$, where $D=\sqrt{x_{\rm o}^{2}+y_{\rm o}^{2}+z_{\rm o}^{2}}$ is the distance between two components of the binary system versus time. When the source star is passing behind the compact component, this phase angle changes from $0$ to $90^{\circ}$, and when the source star is passing in front of the compact component this phase angle alters from $90$ to $180^{\circ}$. Hence, in the simulation when $\Phi < 90^{\circ}$ (or $x_{\rm o}<0$), we calculate the lensing effect. 

\noindent The lensing magnification factor depends on the source radius projected on the lens plane and normalized to the Einstein radius $\rho_{\star}$, and the lens-source distance projected on the lens plane and normalized to the Einstein radius $u$. In our formalism, these two parameters are given by 
\begin{eqnarray}
\rho_{\star}&=&\frac{R_{\star} D_{\rm l}}{R_{\rm E} (D_{\rm l}-x_{\rm o})},~~~ R_{\rm E}=\sqrt{\frac{4 G M_{\rm c}}{c^{2}}\frac{D_{\rm l}~|x_{\rm o}|}{D_{\rm l}-x_{\rm o}}},\nonumber\\
u&=&\frac{d_{\rm p}}{R_{\rm E}},~~~ d_{\rm p}= \sqrt{y_{\rm o}^{2}+z_{\rm o}^{2}},
\label{re}
\end{eqnarray} 
where, $D_{\rm l}$ is the distance of the compact object from the observer, $R_{\star}$ is the radius of the luminous star, $G$ is the gravitational constant, $c$ is the light speed, and $d_{\rm p}$ is the projected distance between the compact object and the source star on the sky plane. 

\noindent Another factor which impacts the lensing magnification is the limb-darkening effect for stellar surface brightness. In the simulation, we consider a linear limb-darkening profile for the source star, as $I=I_{0}\big[1-\Gamma (1-\mu)\big]$, where $I_{0}$ is the stellar brightness at the center of the source disk, $\Gamma$ is the so-called limb-darkening coefficient, and $\mu=\sqrt{1-R^{2}\big/R_{\star}^{2}}$, where $R$ is the radial distance over the source disk. We determine the magnification factor, $A\big(u,~\rho_{\star},~\Gamma\big)$, using the public RT-model \citep{2010Bozza,2018Bozza}. We ignore the General Relativistic (GR) effects on the magnification factor because this effect is significant only for a very small part of the source disk which is exactly behind and collinear with the compact object as seen by the observer. 

In self-lensing events, the Einstein radius (Eq. \ref{re}) is estimated using $R_{\rm E} \simeq \sqrt{4G M_{\rm c} a}/c$, which is considerably smaller than the Einstein radius for common microlensing events toward the Galactic bulge. Hence, in self-lensing events, $\rho_{\star}\gtrsim1$ and their magnification factors are estimated by the ratio of the area of the images' ring generated at the time of the complete alignment to the source area \citep[e.g.,  ][]{2016ApJHan}. The inner and outer radii of the images' ring are (respectively): 
\begin{eqnarray}
R_{\rm{in}}&=& \frac{1}{2}\Big[\sqrt{R_{\star, \rm p}^{2}+4 R_{\rm E}^{2}}-R_{\star, \rm p}\Big],\nonumber \\
R_{\rm{out}}&=& \frac{1}{2}\Big[\sqrt{R_{\star, \rm p}^{2}+4 R_{\rm E}^{2}}+R_{\star, \rm p}\Big], 
\end{eqnarray}    
where, $R_{\star, \rm p}=R_{\star}D_{\rm l}\big/(D_{\rm l}-x_{\rm o})$ is the projected source radius on the lens plane. Accordingly, the magnification factor for a uniform source star during a self-lensing signal is given by \citep{1973AAMaeder,1996Gould,2003ApJEric}: 
\begin{eqnarray}
A\simeq \frac{R_{\rm{out}}^{2}-R_{\rm{in}}^{2}}{R_{\star, \rm p}^{2}}\simeq 1+ \frac{2}{\rho_{\star}^{2}}. 
\end{eqnarray}

We note that when $x_{\rm o}<0$ and during self-lensing signals, the compact object (the gravitational lens) can block some part of images' area, which is the so-called finite-lens effect or occultation of images' light by the compact object \citep[see, e.g.,][]{2001MNRASMarsh,2016ApJHan}. Finite-lens effect is considerable in WDMS binaries. This effect decreases the magnification factor by $\mathcal{O}=R_{\rm c}^{2}\big/R_{\star, \rm p}^{2}$ when $R_{\rm in}\leq R_{\rm c}\leq R_{\rm out}$, and  does not change the magnification factor ($\mathcal{O}=0$) if $R_{\rm c} < R_{\rm in}$. Here, $R_{\rm c}$ is the radius of the compact object. If the radius of the compact object is larger than the outer radius of the images' ring, the light of images is completely blocked by the compact object and $\mathcal{O}=A$.

Whenever $x_{\rm o}>0$ (or $\Phi\in [90,~180^{\circ}]$) the source star is passing in front of its compact companion and it can block its luminosity, the so-called eclipsing effect (when the compact objects are either WDs or NSs with the masses $M_{\rm c}\lesssim 2.9 M_{\odot}$). According to our formalism, eclipsing features happen when $x_{\rm o}>0$ and $d_{\rm p}\leq\big(R_{\rm c}+R_{\star, \rm p}\big)$. We calculate the fraction of the compact object's disk eclipsed by the stellar companion, $\mathcal{E}(x_{\rm o}, y_{\rm o}, z_{\rm o})$, numerically by: 
\begin{eqnarray}
\mathcal{E}(x_{\rm o}, y_{\rm o}, z_{\rm o})= \frac{1}{\pi R_{\rm c}^{2}}\int_{-R_{\rm c}}^{R_{\rm c}} dy'\int_{-\sqrt{R_{\rm c}^{2}-y'^{2}}}^{\sqrt{R_{\rm c}^{2}-y'_{2}}}dz' \Theta \Big[\frac{R_{\star, \rm p}}{d'_{\rm p}}\Big], 
\end{eqnarray}
where $d'_{\rm p}=\sqrt{(y_{\rm o}-y')^2+(z_{\rm o}-z')^2}$ is the distance of each element over the compact object's disk $(y',~z')$ from the source center projected on the sky plane. $\Theta$ is a step function which is one if its argument is larger than one and it is zero when the argument is less than one. We note that the whole disk of the compact object is eclipsed (i.e., $\mathcal{E}=1$) if $d_{\rm p}\leq (R_{\star, \rm p}-R_{\rm c})$, and there is no eclipsing ($\mathcal{E}=0$) when either $x_{\rm o}<0$, or $x_{\rm o}>0$ and $d_{\rm p}\geq(R_{\rm c}+R_{\star, \rm p})$.

The apparent magnitude of the source star by considering self-lensing, occultation (or finite-lens effect), and eclipsing signals in edge-on binary systems as measured by the observer is given by:  
\begin{eqnarray}
m_{\rm{o}}=m_{\star}-2.5 \log_{10}\Big[f_{\rm b}\frac{A(u, \rho_{\star}, \Gamma)-\mathcal{O}+\mathcal{F} \mathcal{E}}{1+\mathcal{F}}+1-f_{\rm b}\Big]
\label{magnitude}
\end{eqnarray}
where, $\mathcal{F}$ is the ratio of the compact object's flux to the stellar flux, and $f_{\rm b}$ is the blending factor which is the ratio of the source flux to the total flux received from the source star PSF (Point Spread Function), $m_{\star}$ is the apparent magnitude of source star when it is isolated and without any companion.  

\begin{figure*}
\centering
\includegraphics[width=0.49\textwidth]{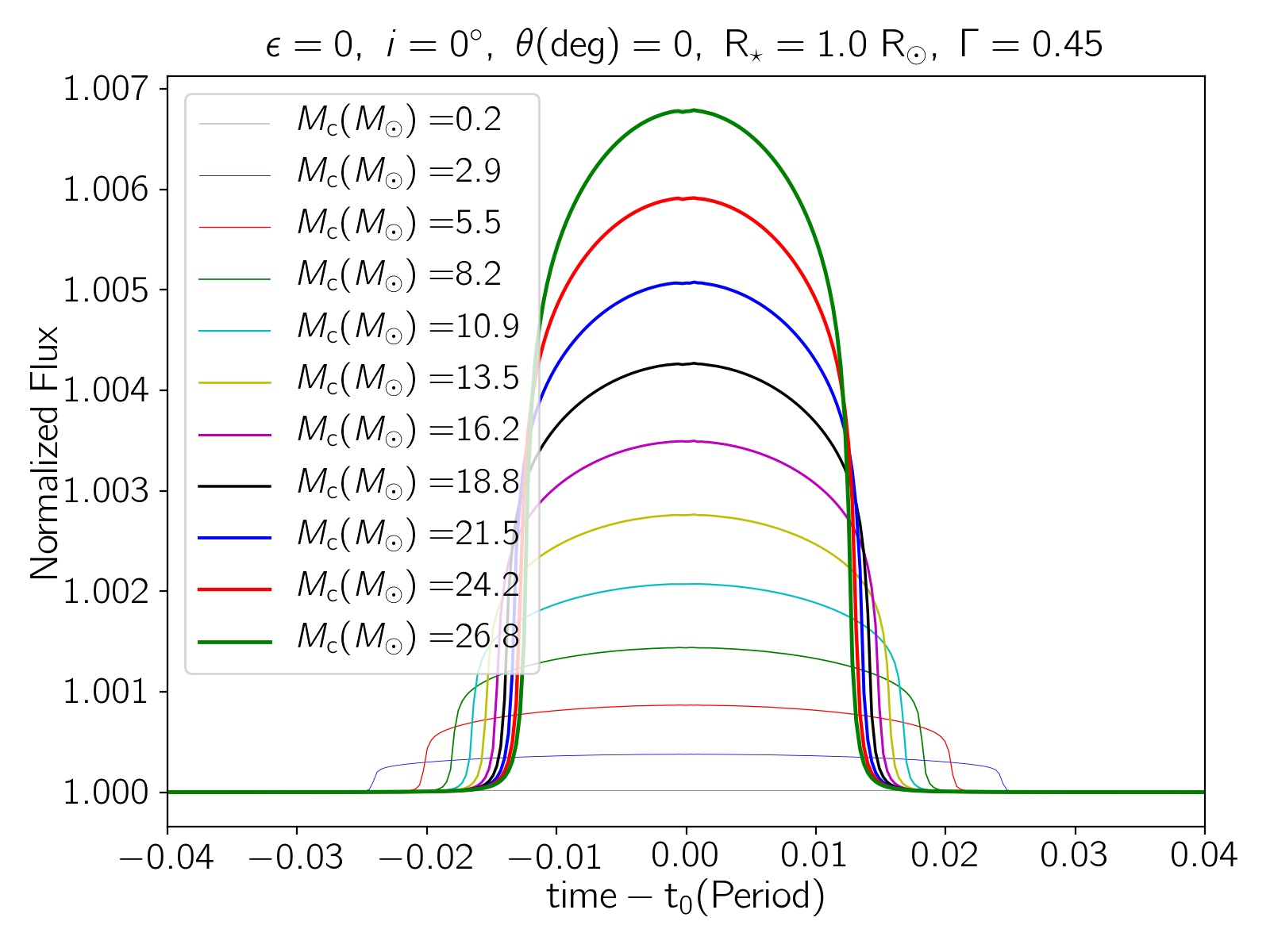}
\includegraphics[width=0.49\textwidth]{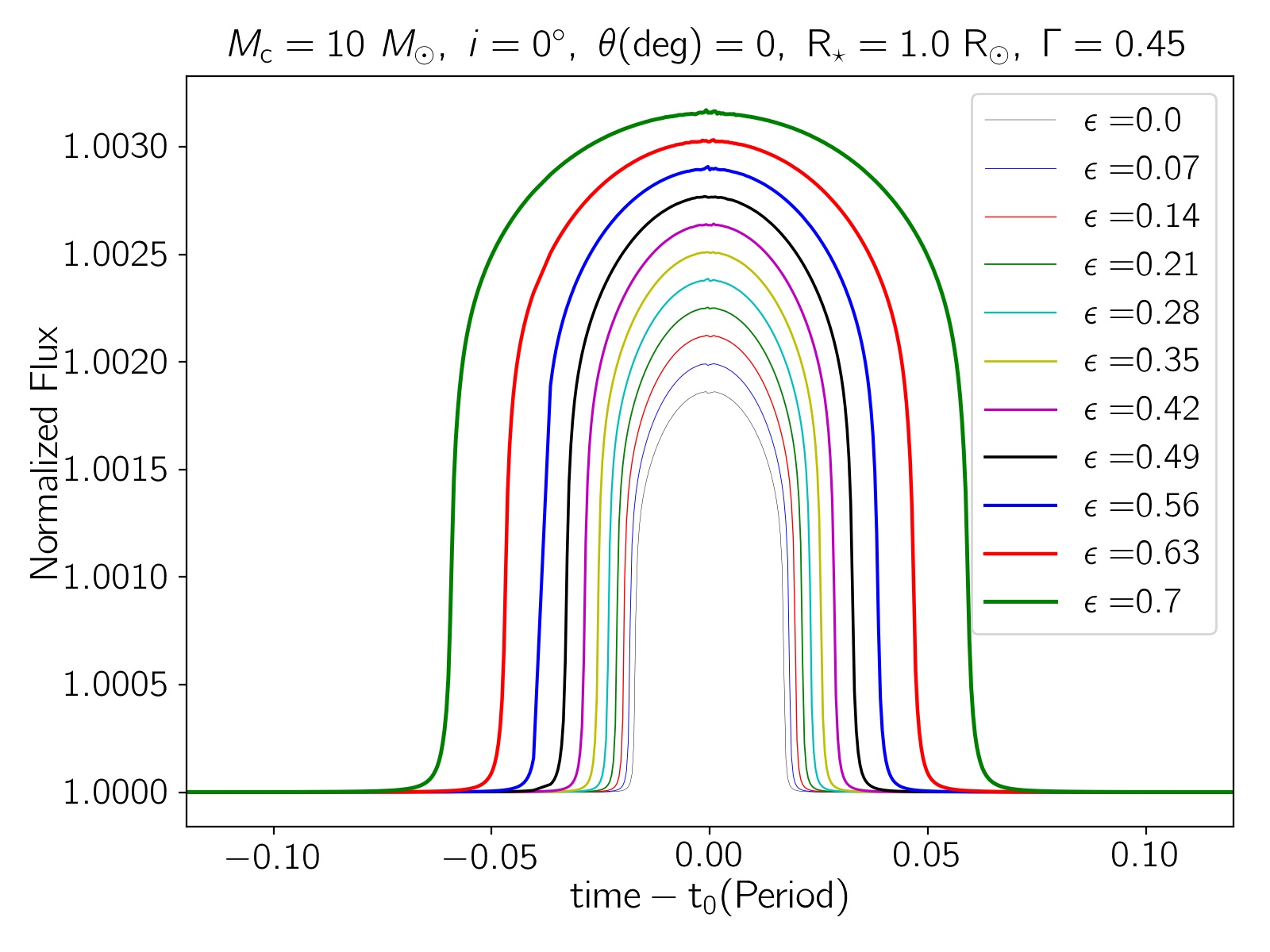}
\includegraphics[width=0.49\textwidth]{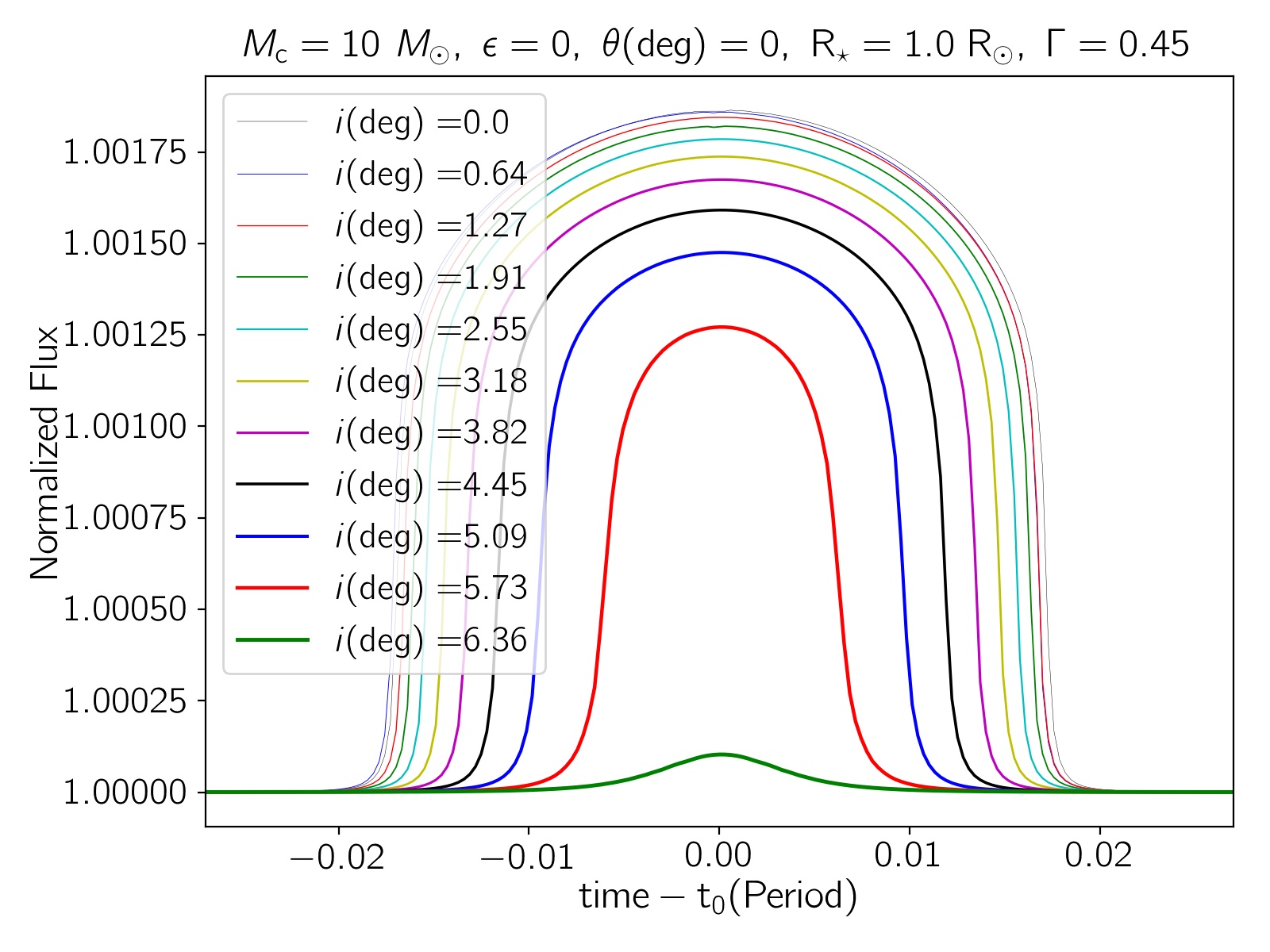}
\includegraphics[width=0.49\textwidth]{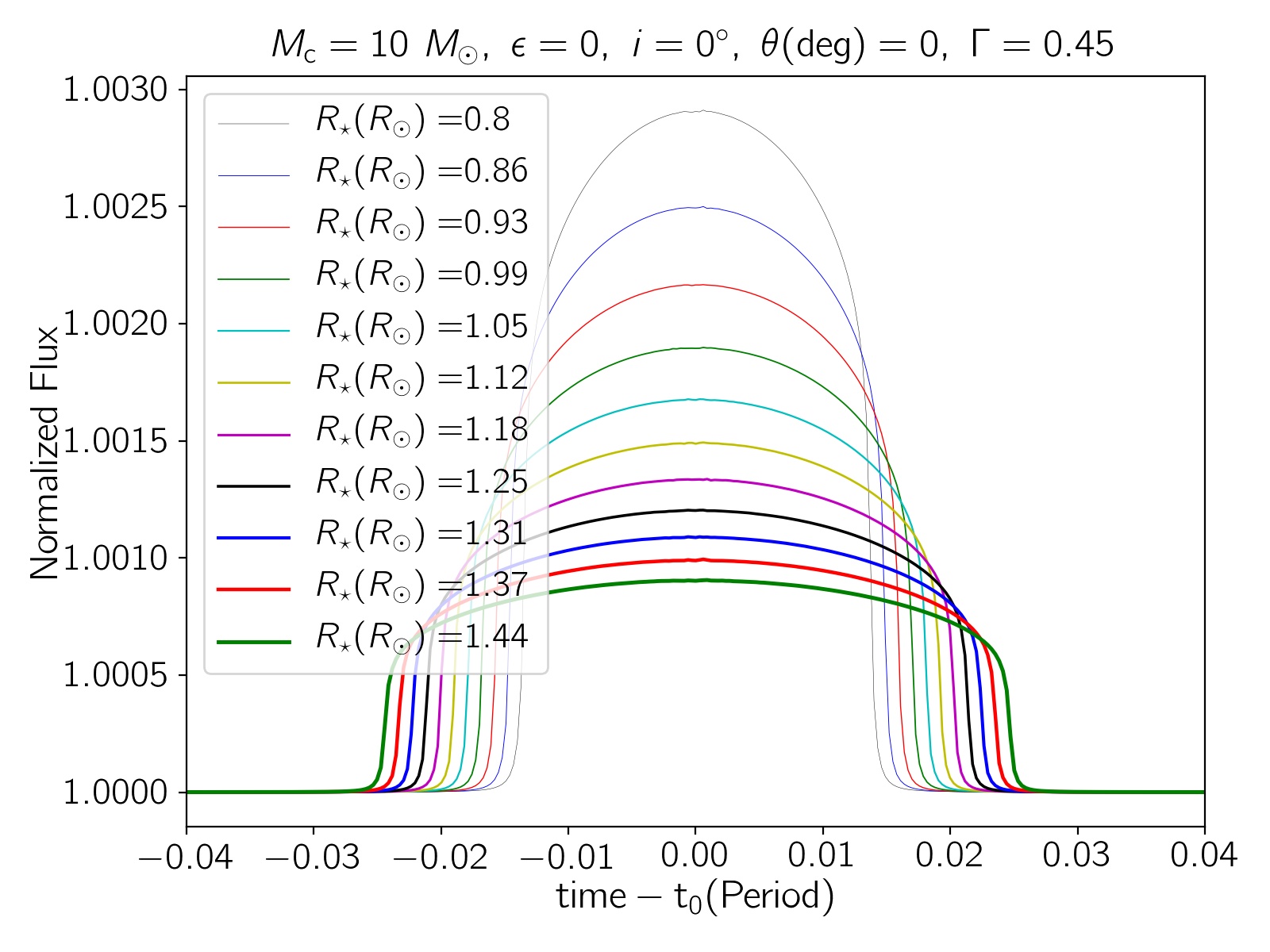}
\includegraphics[width=0.49\textwidth]{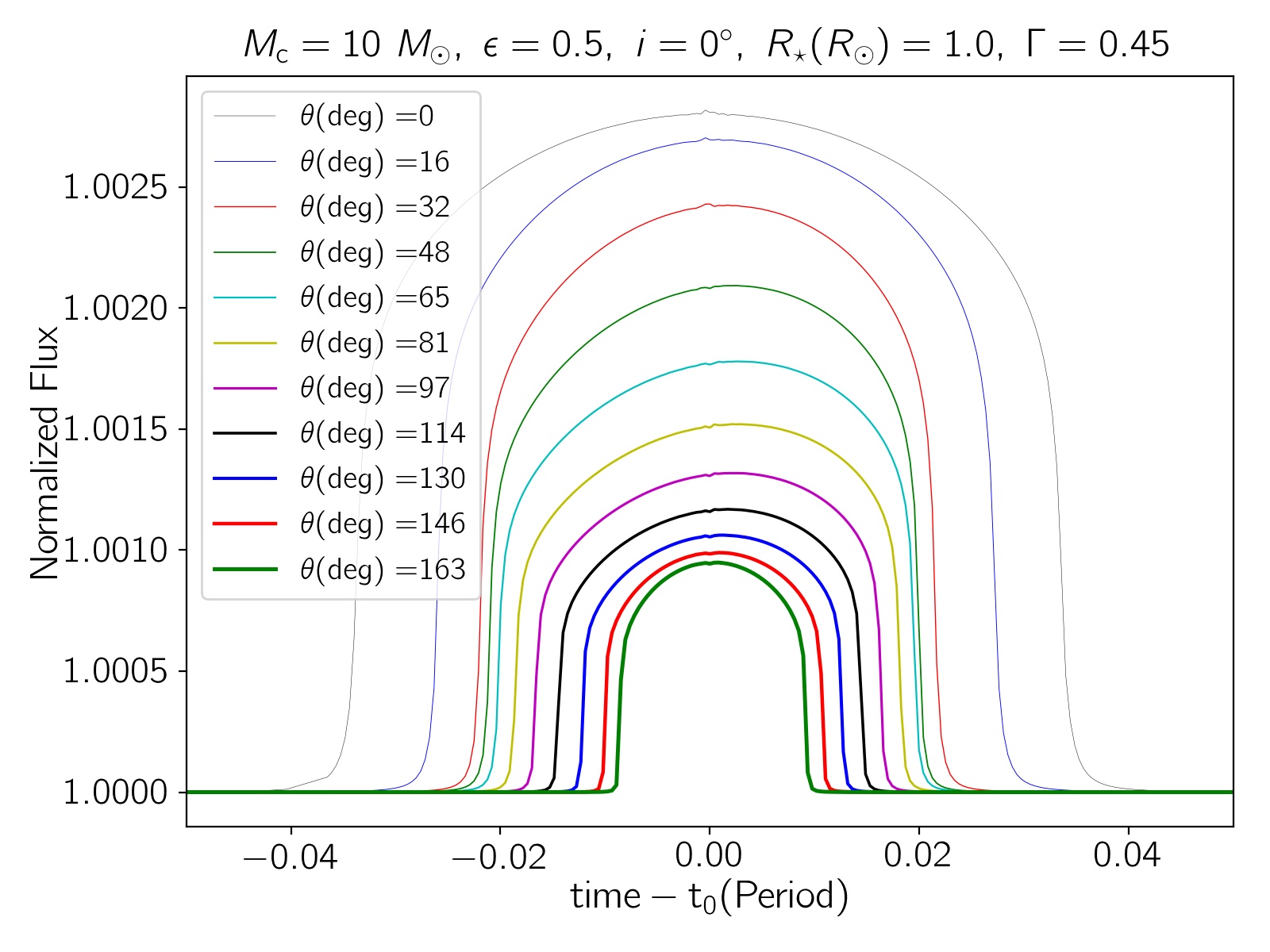}
\includegraphics[width=0.49\textwidth]{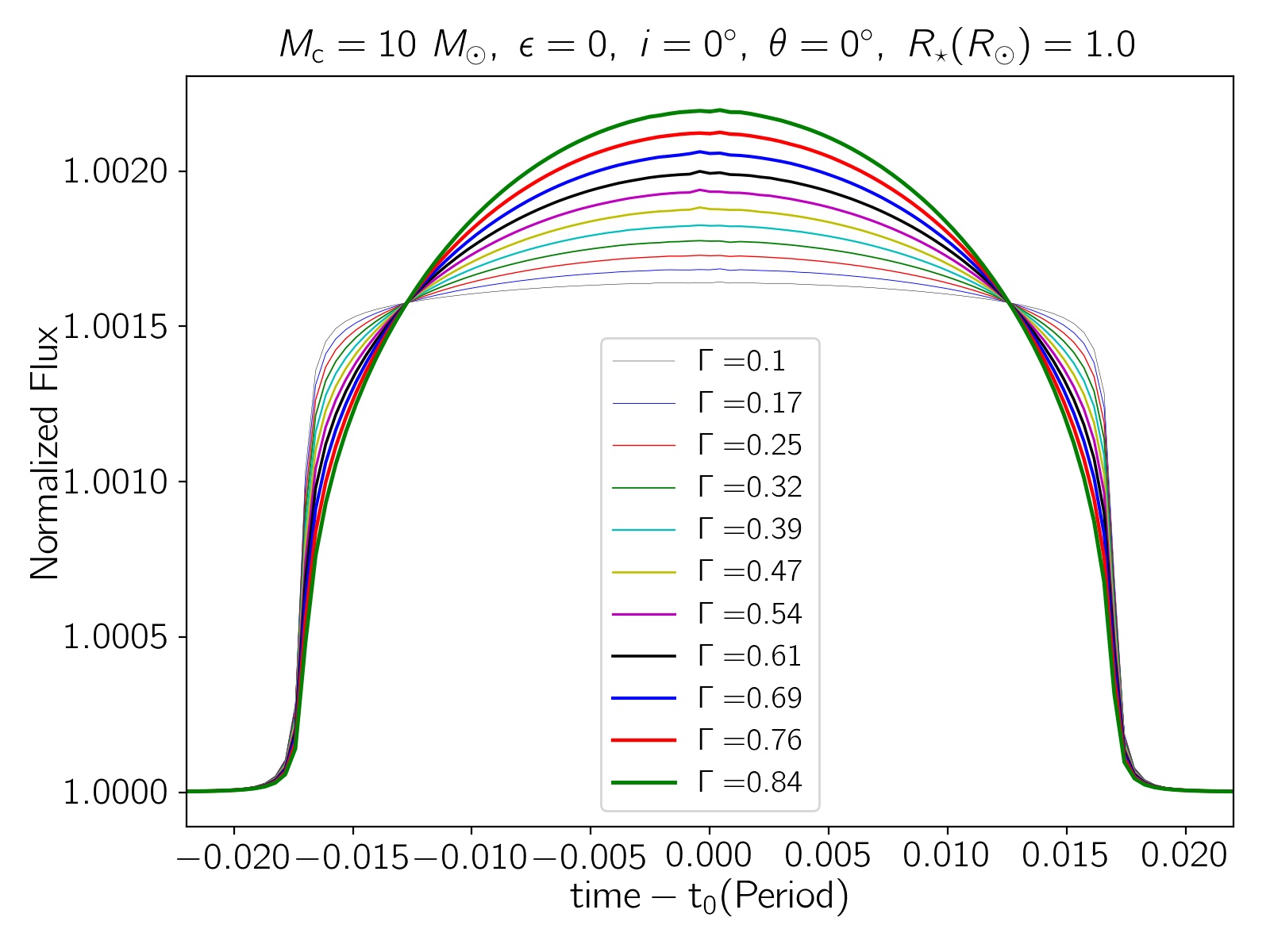}
\caption{Simulated self-lensing signals versus time (normalized to the orbital period) by considering different values for $M_{\rm c}(M_{\odot})$, $\epsilon$, $i$, $R_{\star}(R_{\odot})$, $\theta$, and $\Gamma$. For each panel, several parameters are fixed and reported at the top of that panel. The vertical axis is the flux of the source star normalized to its baseline value.}\label{simu}
\end{figure*}

\subsection{Characteristics of Self-Lensing Signals}\label{sub3}
Here, we evaluate the peak amounts in self-lensing signals, i.e., $\Delta F_{\rm L}\simeq 2~\rho_{\star}^{-2}$, and consider four types of source stars: (i) a red dwarf with $M_{\star}=0.3~M_{\odot}$, and $R_{\star}=0.35~R_{\odot}$, (ii) a Sun-like star, (iii) a B-star with $M_{\star}=2~M_{\odot}$, and $R_{\star}=1.8~R_{\odot}$, and (iv) a sub-giant star with $M_{\star}=2~M_{\odot}$, and $R_{\star}=3.6~R_{\odot}$, as well as two wide ranges for the compact object's mass and the orbital semi-major axis as $M_{\rm c}(M_{\odot})\in [0.1,~50]$, and $a(R_{\sun})\in [5,~5000]$, respectively. For all of these binary systems, we calculate $\Delta F_{\rm L}$ values and show their maps in Figure \ref{maps}.  

Over these maps, the contours of $\log_{10}[T(\rm{days})]$ are shown with black solid and dashed lines (corresponding to positive and negative values, respectively). In these plots, binary systems with source radii larger than the Roche-Lobe distance are excluded and covered with white colour on maps. For these systems, the mass transfers from the bright star to the compact object and there are some other sources for variability (e.g., ellipsoidal effect) in stellar fluxes which makes discerning self-lensing signals hard and impossible. Here, we determine the Roche-Lobe distance from the source center as \citep{1971Paczynski, 1983ApJRoche}: 
\begin{eqnarray}
D_{\rm{RL}}=a-a \frac{0.49~q^{2/3}}{0.6~q^{2/3} +\log_{10}\big[1+q^{1/3}\big]},  
\label{rochelobe}
\end{eqnarray}
where, $q= M_{\rm c}/M_{\star}$. According to these plots, generally more massive compact objects in wider orbits (with larger semi-major axes) have larger lensing-induced signals, so that increasing (a) the mass of compact objects from $0.1 M_{\odot}$ to $50 M_{\odot}$, and (b) the semi-major axis by three orders of magnitude (from $5 R_{\sun}$ to $5000 R_{\sun}$) enlarge $\Delta F_{\rm L}$ by $\sim 5$ orders of magnitude. However, more massive compact objects in wider orbits have longer orbital periods. For that reason, less massive source stars (e.g., red dwarfs in comparison to B-stars) are more suitable for  detecting their self-lensing signals generated by their compact companions because (i) they generate shorter orbital periods, and (ii) they have smaller radii, smaller $\rho_{\star}$ values, and as a result higher $\Delta F_{\rm L}$ values. We probe this point again in the next section and through Monte Carlo simulations. 

Generally, peaks, shapes and durations of self-lensing signals determine their detectability. They depend on several parameters including $M_{\rm c}$, $\epsilon$, $i$, $R_{\star}$, $\theta$, and $\Gamma$. We therefore simulate self-lensing signals by considering different values of these parameters as shown in different panels of Figure \ref{simu}. In each panel, one parameter changes and other parameters are fixed and mentioned at the top of that panel. Accordingly, we summarize some points in follows.  

\begin{itemize}[leftmargin=2.5mm]
\item The maximum enhancement in self-lensing signals increases with the lens mass as $\Delta F_{\rm L} \propto M_{\rm c}$. The self-lensing signals due to completely edge-on orbits are flattened (top-hat models).

\item There is a degeneracy between $M_{\rm c}$ and $R_{\star}$, so that small stellar radii make similar self-lensing signals to those due to more massive compact objects. Both of these parameters change the peaks of self-lensing from top-hat ones (but they do not change the self-lensing edges). 

\item If the stellar orbits are eccentric, the resulting self-lensing signals are asymmetric, unless the source star is passing from either periapsis or apoapsis point of its orbit while lensing (which are rare). This point can be found from the fifth panel of Figure \ref{simu}.   

\item The inclination angle of stellar orbit with respect to the line of sight is the only factor which causes that self-lensing signals at the edges are not broken (not a strict top-hat model). By increasing the inclination angle self-lensing signals at edges are rather slow-enhancing.  

\item In more eccentric stellar orbits, the resulting self-lensing signals could be wider depending on the source-lens distance while lensing (in our formalism this distance is determined by $\theta$). 

\item The limb-darkening effect has a very small impact on the width of self-lensing signals, whereas it changes the peak and shape of self-lensing signals.
\end{itemize}   

In the next step, we perform Monte Carlo simulations from all possible self-lensing signals due to different binary systems (including main-sequence stars and compact objects) and study their detectability in the TESS data. 
\begin{figure*}
\centering
\includegraphics[width=0.49\textwidth]{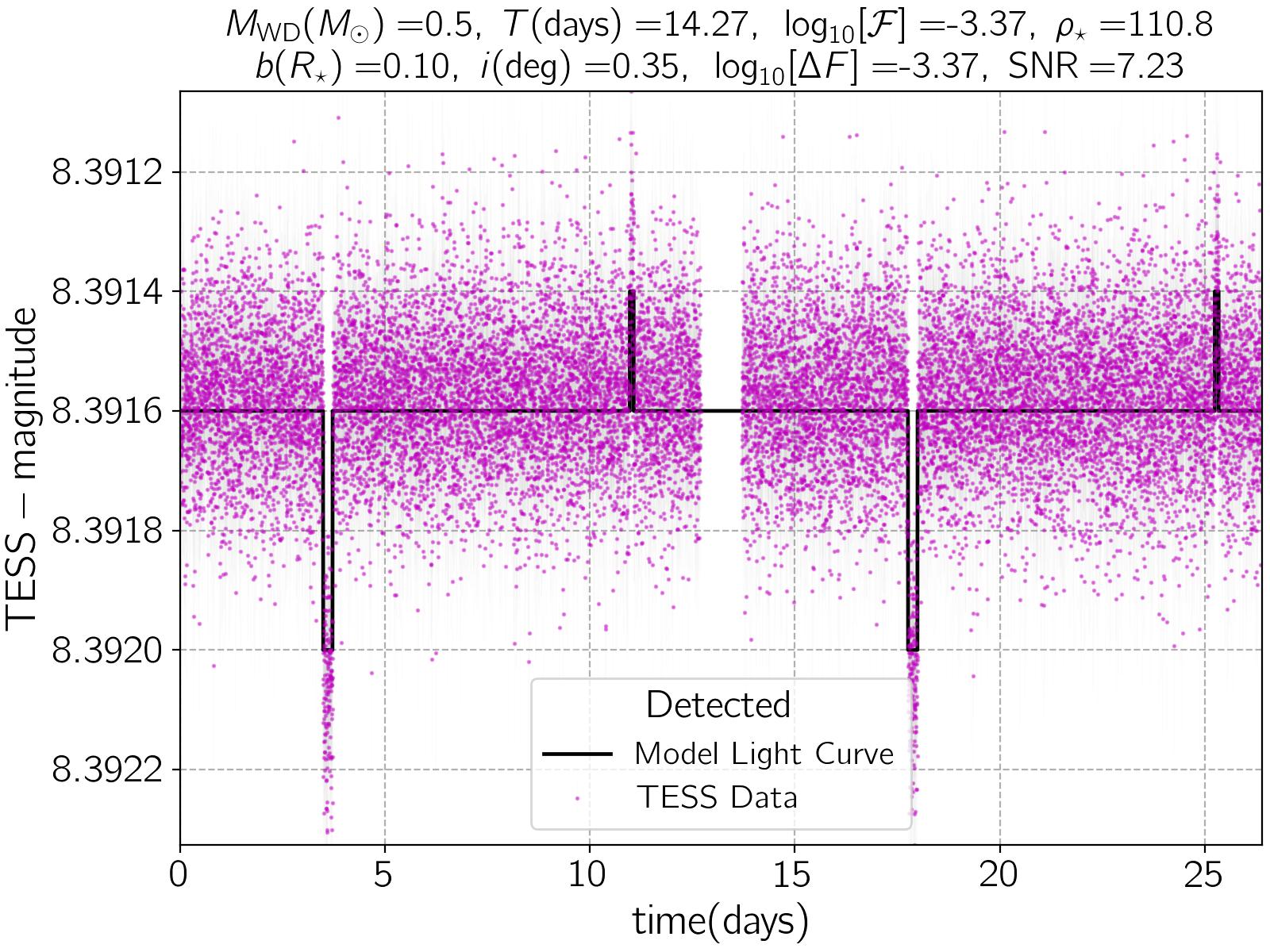}
\includegraphics[width=0.49\textwidth]{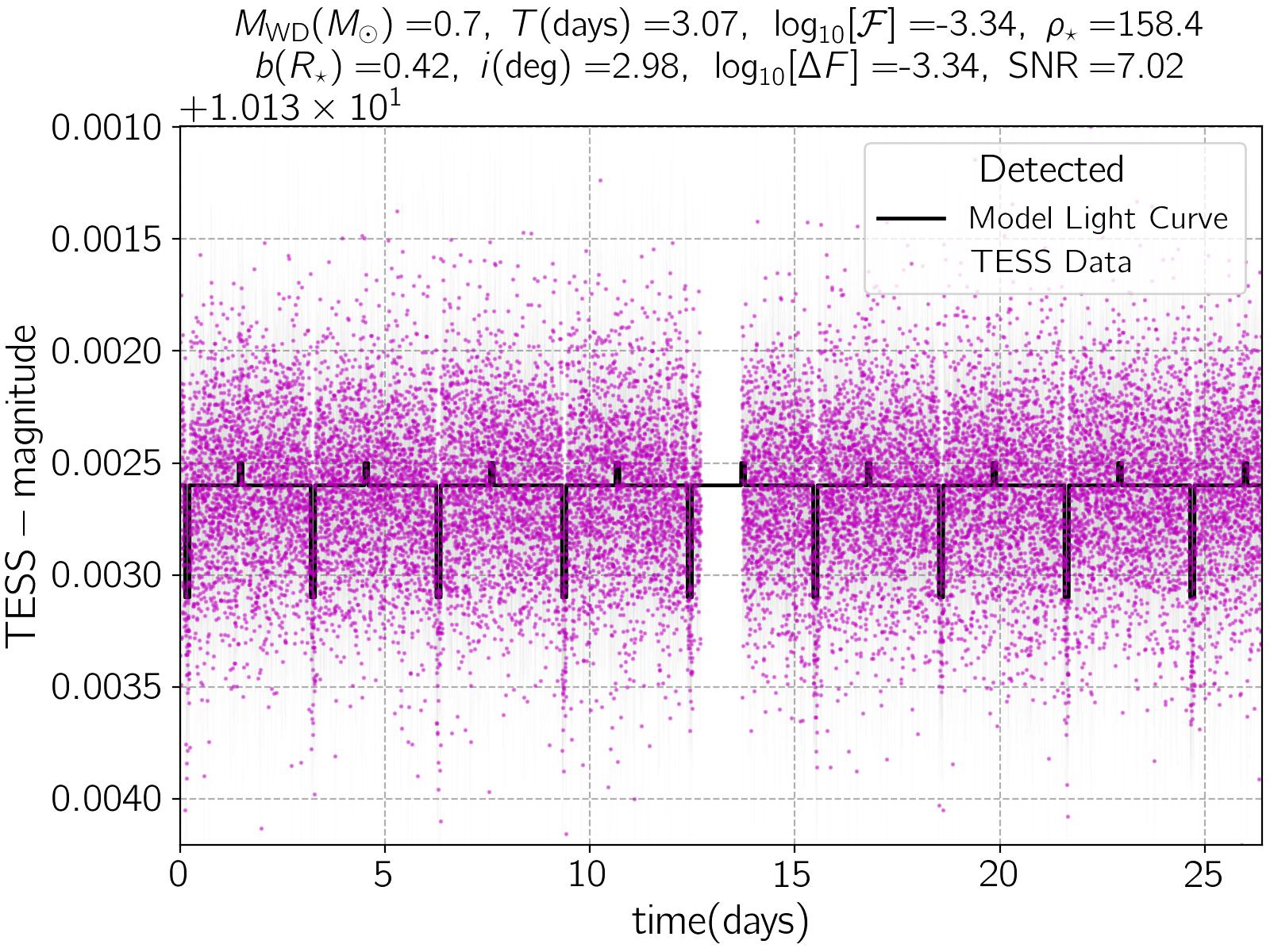}	
\includegraphics[width=0.49\textwidth]{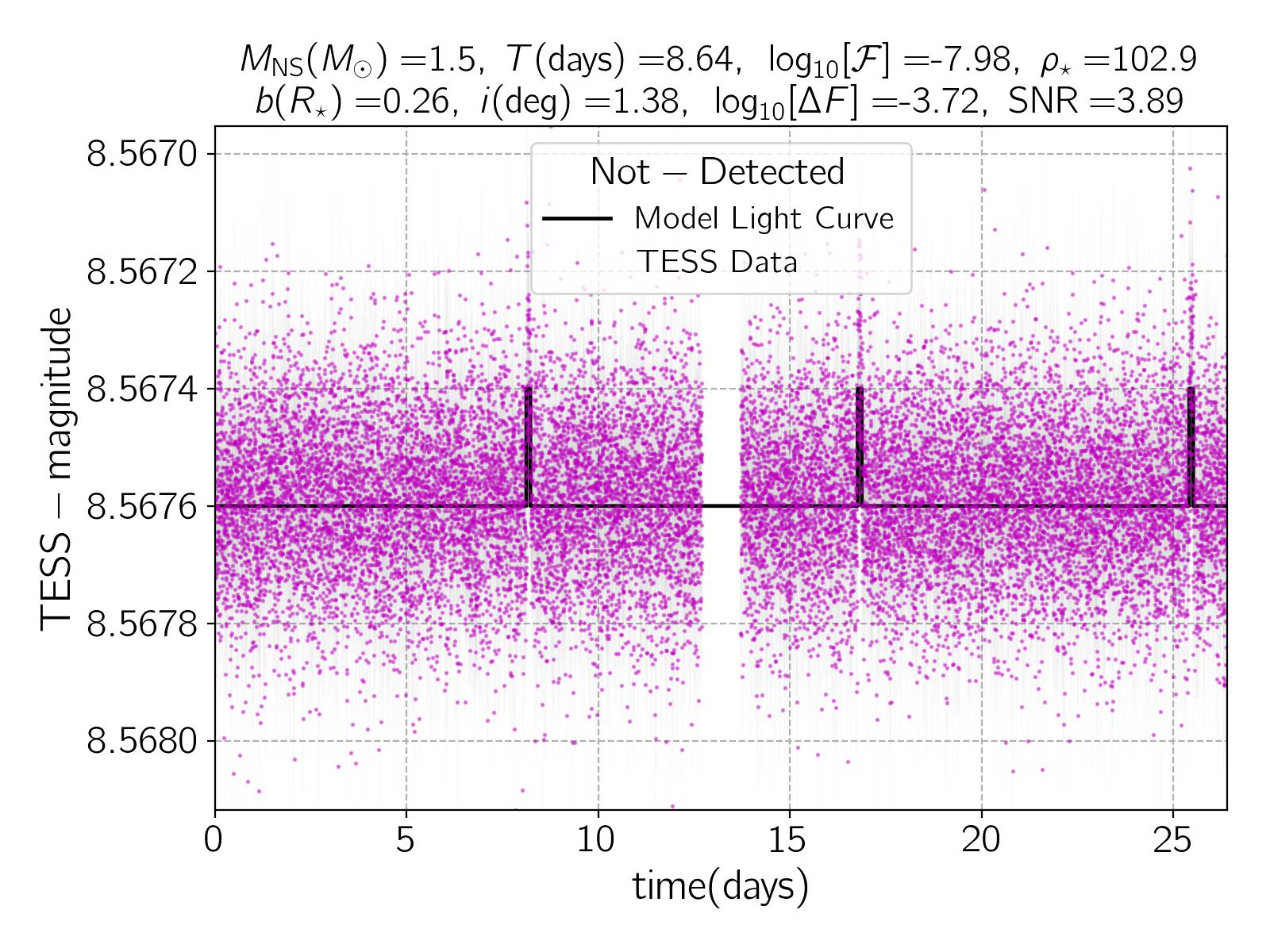}
\includegraphics[width=0.49\textwidth]{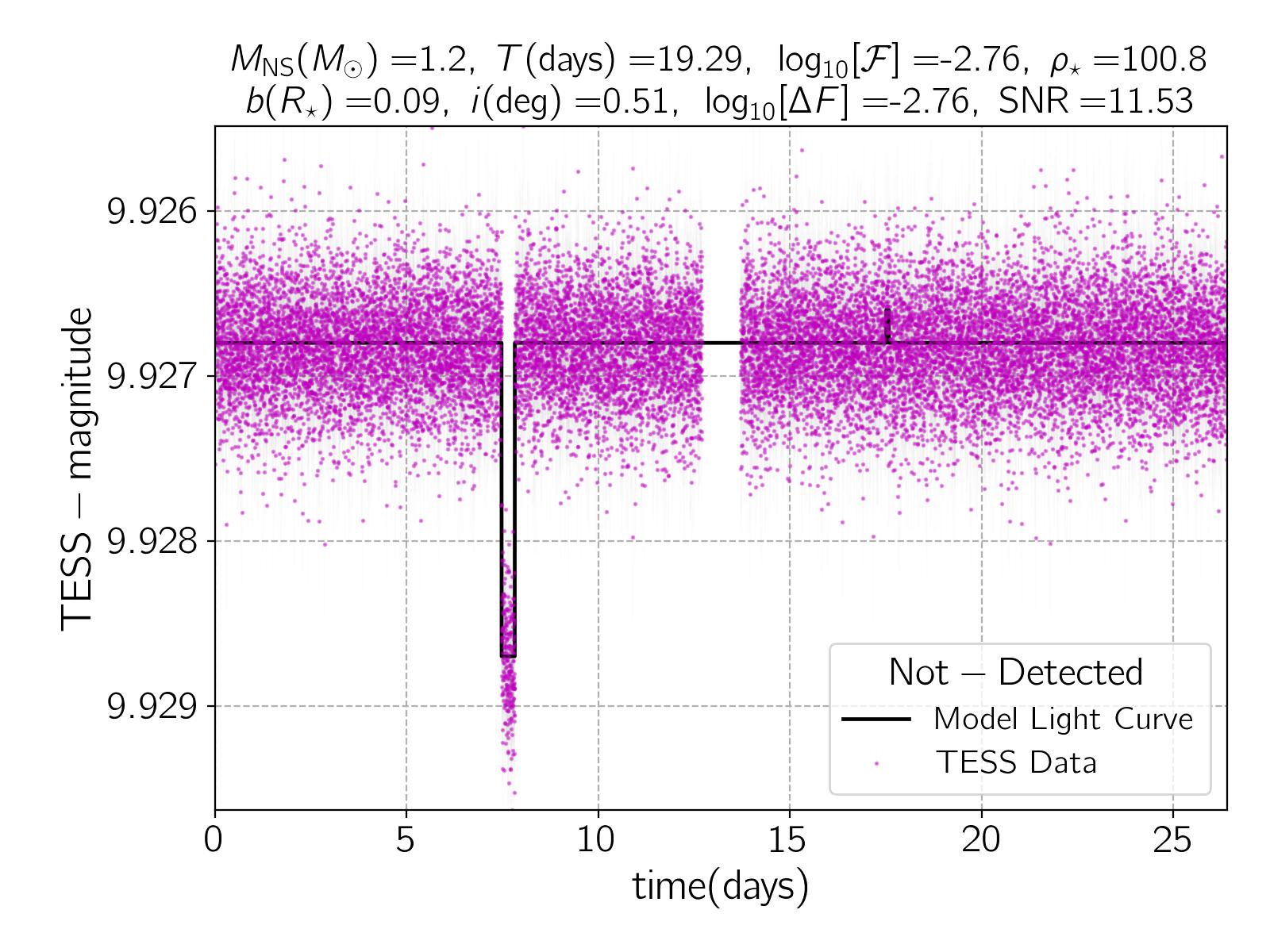}
\includegraphics[width=0.49\textwidth]{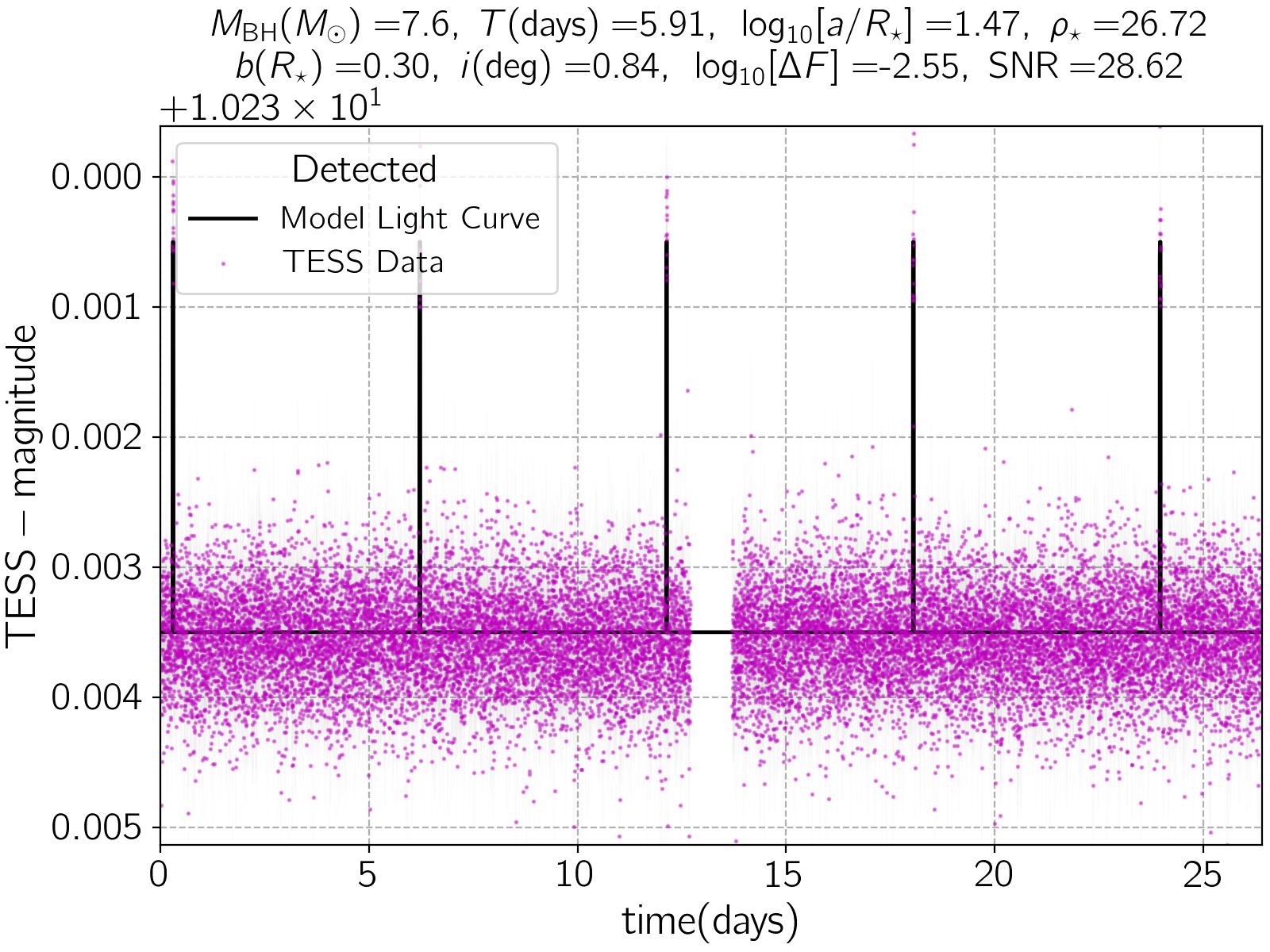}
\includegraphics[width=0.49\textwidth]{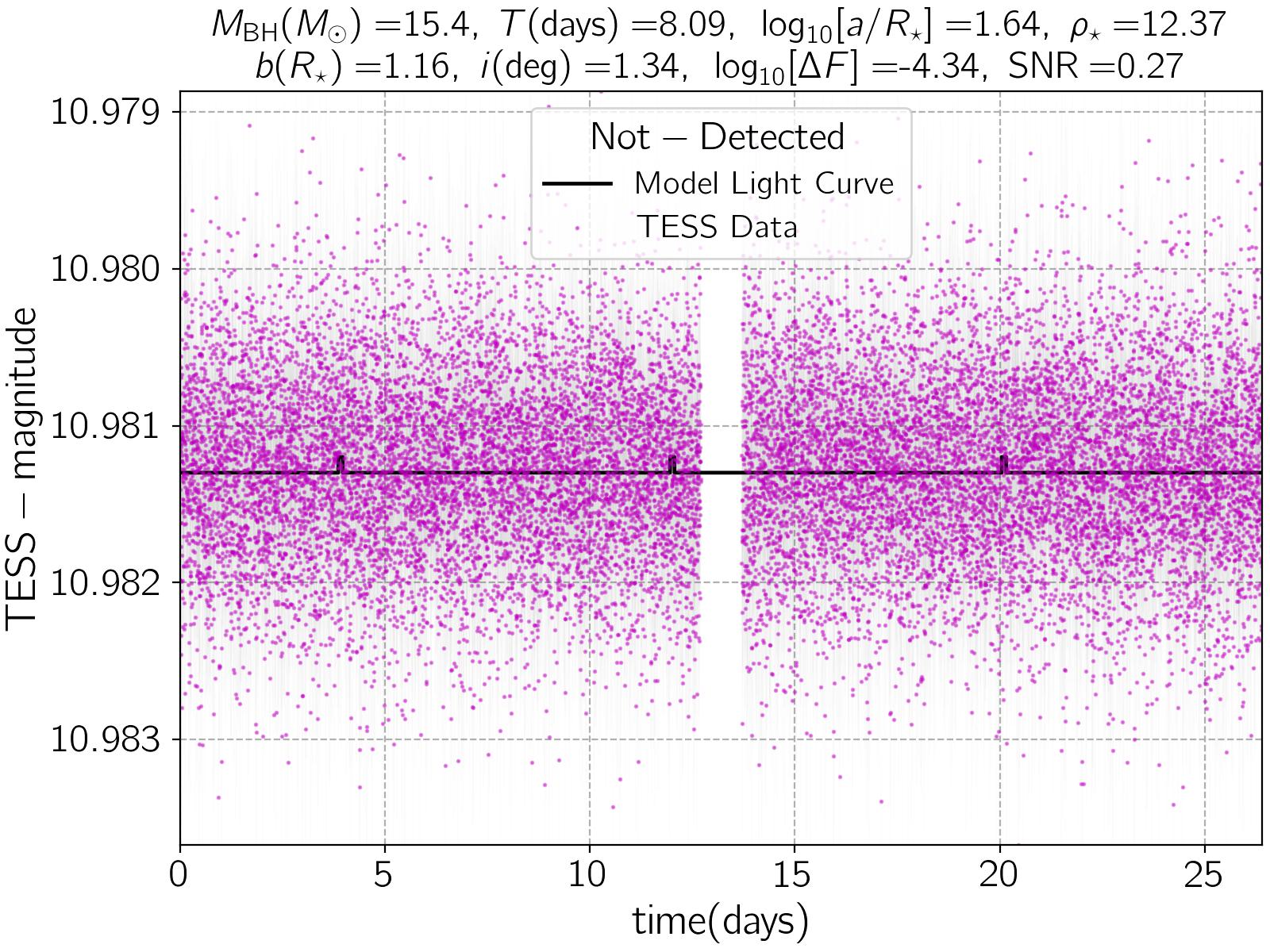}
\caption{Six examples of simulated stellar light curves due to detached and edge-on binary WDMS, NSMS, and BHMS systems. The synthetic data points are hypothetically taken by the TESS telescope. We fix the observational time to $27.4$ days with a one-day gap in middle. The parameters used to make light curves are mentioned at the tops of panels. By applying HC detectability criteria, for three of these light curves the impacts of compact objects are realizable, as mentioned in their legends.}\label{self}
\end{figure*}

\section{Monte Carlo Simulations}\label{sec2}
To simulate potential self-lensing, occultation, and eclipsing signals that can be detected in the TESS observations, we first take an ensemble of the Candidate Target List (CTL, \citet{2018AJStassun,2019AJStassun}) from the Mikulski Archive for Space Telescopes (MAST) catalog \citep{ctlTESS}. The TESS CTL targets are relatively close and bright stars with pre-measured physical parameters which were (and are) observed by the TESS telescope with a $2$-min cadence, based on their priority. In this ensemble, for CTL targets their mass $M_{\star}$, priority, blending factor $f_{\rm b}$, distance $D_{\rm l}$, source radius $R_{\star}$, effective surface temperature $T_{\star}$, and stellar apparent magnitude in the TESS filter $m_{\star}$ are reported which all are used for simulating source stars in binary systems.  

We assume the TESS CTL targets live in detached binaries with compact companions. We take $1772$ known WDs which were extracted from the SDSS data with the reported mass $M_{\rm{WD}}$, distance, radius $R_{\rm{WD}}$, and apparent magnitudes in the Gaia passbands, i.e., $m_{G}$, $m_{G_{\rm{BP}}}$, and $m_{G_{\rm{RP}}}$ at the distances closer than $100$ parsec from \citet{2020ApJkilic}. Using the same method which was explained in \citet{sajadian2024}, we convert their apparent magnitudes in the Gaia filters to their absolute magnitudes in the TESS passband.

The numbers of discovered NSs and SBHs up to now are low. Additionally, all necessary parameters for discovered ones were not measured. Therefore, in Monte Carlo simulations we generate their populations synthetically according to the known distribution functions for their physical parameters. 

The mass distribution function of NSs is a bimodal distribution with two peaks at $1.37 M_{\odot}$, and $1.73 M_{\odot}$ whereas the second one is wider and more flattened \citep{NSmassDis}. So, we determine the mass of NSs using $\mathcal{N}\big(1.37,~0.15\big)+0.5~ \mathcal{N}\big(1.73,~0.3\big)$ from the range $ M_{\rm{NS}}\in [0.5,~2.9] M_{\odot}$, where $\mathcal{N}(\mu,~\sigma)$ is a normal function with the mean value $\mu$, and the width $\sigma$. The radius of NSs is a function of their mass and is in the range $R_{\rm{NS}}\in [10,~15]$ km \citep{NSmassradius}. The luminosity of NSs is a function of their ages and masses, because these objects cool down over time through thermal radiation. We determine the NSs' luminosity based on Figure (2) of \citet{2020Potekhim}. We choose the NSs' ages $(\mathcal{A})$ in the logarithmic scale uniformly from the range $\log_{10}[\mathcal{A}(\rm{year})]\in [2.5,~8]$. For SBHs we choose their masses uniformly from the range $M_{\rm{BH}}\in [3.3,~50]~M_{\odot}$ \citep{2022ApJsicilia,2023AJsajadian}.

In each step of Monte Carlo simulations, we take one CTL target and one compact object as two components of a detached binary system. We take the semi-major axis of their orbit $a$ from a log-uniform distribution and in the range $[3R_{\star},~10^{6}R_{\star}]$ \citep[see, e.g., ][]{1983Abt}, and put aside the interacting systems with the source radii larger than the Roche-Lobe distance ($D_{\rm{RL}}$ which is given by Eq. \ref{rochelobe}). 

The known period-eccentricity correlation indicates an upper limit on the orbital eccentricity $(\epsilon_{\rm{max}})$ for a given orbital period $T$ \citep{2008EASMazeh}. By considering this correlation, we uniformly choose the orbital eccentricity from the range $\epsilon \in [0, \epsilon_{\rm{max}}]$. We choose the inclination angle uniformly from the range $i\in [0,~20^{\circ}]$, because for larger inclination angles neither self-lensing/occultation nor eclipsing signals happen. We choose the projection angle $\theta$ uniformly from the range $[0,~360^{\circ}]$.  

In the next step, we generate synthetic data points taken by the TESS telescope. We fix the observing cadence to $2$ minutes, and consider a one-day gap after each $13.7$-day observation. The observing time span is calculated by $T_{\rm{obs}}=\rm{No}_{\rm s}\times27.4$ days, where $\rm{No}_{\rm s}$ is the number of overlapping sectors and can be an integer number from one to thirteen. Therefore, the longest observing time is $\sim 360$ days for the ecliptic poles during one year.

To examine their detectability we calculate the signal-to-noise ratio (SNR), which is defined for planetary transit events, as given by \citep[see, e.g., ][]{2024Fatheddin}: 
\begin{eqnarray}
\rm{SNR}= \sqrt{N_{\rm{tran}}}~\frac{\Delta F \times 10^{6}}{\rm{CDPP}},
\label{snr}
\end{eqnarray}
where, CDPP is the Combined Differential Photometric Precision (CDPP) metric which is a function of the stellar apparent magnitude in the TESS passband, $N_{\rm{tran}}=T_{\rm{obs}}\big/T$ is the number of orbital period during an observing time span. $\Delta F$ is the maximum variation in the stellar flux which is due to either self-lensing/occultation or eclipsing signals (the maximum of $\Delta F_{\rm L}$, $\Delta F_{\rm O}$, and $\Delta F_{\rm E}$). In the simulation, to extract the detectable events we consider two sets of criteria which are (i) $\rm{SNR}>5$, and $N_{\rm{tran}}>2$, and (ii) $\rm{SNR}>3$, and $N_{\rm{tran}}>1$. The first set extracts detectable events with a high confidence (HC), and the second one takes detectable events with a low confidence(LC).

In Figure \ref{self}, we show six examples of simulated stellar light curves due to edge-on WDMS, NSMS, BHMS systems (with $i\leq20^{\circ}$). Some of useful parameters to make light curves and their SNR values are reported at tops of plots. The magenta synthetic data points are taken by the TESS telescope with a $2$-min cadence. Accordingly, three events are detectable (with a high confidence) in the TESS data and three others are not detectable.

\begin{deluxetable*}{c c c c c c c c c c c c c c c}
\tablecolumns{15}
\centering
\tablewidth{0.99\textwidth}\tabletypesize\footnotesize 
\tablecaption{The average parameters of the simulated WDMS binaries which have detectable WD-induced signals with a high confidence in their stellar light curves by considering different observing time spans.\label{tab1}}
\tablehead{\colhead{$\rm{No}_{\rm s}$} & $T_{\rm{obs}}$&\colhead{$\overline{M_{\rm WD}}$}&\colhead{$\overline{T}$} & \colhead{$\overline{\log_{10}[a/R_{\star}]}$} & $\overline{\epsilon}$ & $\overline{i}$ & $\overline{\log_{10}[\mathcal{F}]}$& $\overline{D_{\rm l}}$ & $\overline{\rm{SNR}}$ & $\overline{\log_{10}[\rho_{\star}]}$& $\overline{b}$ & $f_{\rm L}:f_{\rm E}:f_{\rm O}$&\colhead{$\varepsilon_{\rm{HC}}$}&\colhead{$\varepsilon_{\rm{LC}}$}\\
        & $(\rm{days})$ &$(M_{\sun})$&$(\rm{days})$&~&~&$(\rm{deg})$& & $(\rm{kpc})$& ~&~& $R_{\star}$&$[\%]$& $[\%]$&$[\%]$}
    \startdata
$1$ & $27.4$ & $0.65$ & $1.46$ & $0.77$ & $0.04$ & $5.9$ & $-2.66$ & $0.10$ & $195$ & $2.17$ & $0.54$ &  $0:99:1$ & $4.59$&$5.56$\\
$2$ & $54.8$ & $0.65$ & $2.04$ & $0.80$ & $0.06$ & $5.7$ & $-2.78$ & $0.11$ & $238$ & $2.18$ & $0.54$ &  $0:98:2$ & $5.34$&$6.24$\\
$3$ & $82.2$ & $0.65$ & $2.28$ & $0.80$ & $0.06$ & $5.6$ & $-2.83$ & $0.12$ & $276$ & $2.18$ & $0.54$ &  $0:98:2$ & $5.65$&$6.62$\\
$4$ & $109.6$ & $0.65$ & $2.76$ & $0.82$ & $0.07$ & $5.6$ & $-2.86$ & $0.12$ & $305$ & $2.17$ & $0.54$ &  $1:97:3$ & $5.92$&$6.89$\\
$5$ & $137.0$ & $0.65$ & $3.16$ & $0.83$ & $0.07$ & $5.5$ & $-2.90$ & $0.12$ & $326$ & $2.17$ & $0.55$ &  $1:96:3$ & $6.20$&$7.09$\\
$6$ & $164.4$ & $0.65$ & $3.39$ & $0.83$ & $0.07$ & $5.5$ & $-2.92$ & $0.12$ & $348$ & $2.17$ & $0.55$ &  $1:96:3$ & $6.36$&$7.22$\\
$7$ & $191.8$ & $0.65$ & $3.62$ & $0.84$ & $0.08$ & $5.4$ & $-2.93$ & $0.12$ & $368$ & $2.17$ & $0.56$ &  $1:96:3$ & $6.50$&$7.34$\\
$8$ & $219.2$ & $0.65$ & $4.01$ & $0.85$ & $0.08$ & $5.4$ & $-2.95$ & $0.13$ & $385$ & $2.17$ & $0.56$ &  $1:95:3$ & $6.64$&$7.41$\\
$9$ & $246.6$ & $0.65$ & $4.30$ & $0.85$ & $0.09$ & $5.4$ & $-2.96$ & $0.13$ & $403$ & $2.17$ & $0.56$ &  $1:95:3$ & $6.75$&$7.52$\\
$10$ & $274.0$ & $0.65$ & $4.69$ & $0.86$ & $0.09$ & $5.3$ & $-2.97$ & $0.13$ & $418$ & $2.16$ & $0.56$ &  $2:95:4$ & $6.85$&$7.61$\\
$11$ & $301.4$ & $0.65$ & $4.88$ & $0.86$ & $0.09$ & $5.3$ & $-2.98$ & $0.13$ & $434$ & $2.17$ & $0.56$ &  $2:95:4$ & $6.92$&$7.66$\\
$12$ & $328.8$ & $0.66$ & $5.01$ & $0.86$ & $0.09$ & $5.3$ & $-2.99$ & $0.13$ & $451$ & $2.16$ & $0.56$ &  $2:95:4$ & $6.96$&$7.69$\\
$13$ & $356.2$ & $0.66$ & $5.04$ & $0.86$ & $0.09$ & $5.3$ & $-2.99$ & $0.13$ & $465$ & $2.16$ & $0.57$ &  $2:94:4$ & $7.03$&$7.75$\\
\enddata
\tablecomments{$No_{\rm s}$ is the number of overlapping sectors. $\varepsilon_{\rm{HC}}$, and $\varepsilon_{\rm{LC}}$ are the efficiencies for detecting WD-induced signals with high and low confidences, respectively.}
\end{deluxetable*}
\begin{deluxetable*}{c c c c c c  c c c c c c c c c}
    \tablecolumns{15}
    \centering
    \tablewidth{0.99\textwidth}\tabletypesize\footnotesize
    \tablecaption{Same as Table \ref{tab1}, but for simulated NSMS binaries.\label{tab2}}
    \tablehead{\colhead{$\rm{No}_{\rm s}$} & $T_{\rm{obs}}$&\colhead{$\overline{M_{\rm{NS}}}$}&\colhead{$\overline{T}$} & \colhead{$\overline{\log_{10}[a/R_{\star}]}$} & $\overline{\epsilon}$ & $\overline{i}$ & $\overline{\log_{10}[\mathcal{F}]}$& $\overline{D_{\rm l}}$ & $\overline{\rm{SNR}}$ & $\overline{\log_{10}[\rho_{\star}]}$& $\overline{b}$ & $f_{\rm L}:f_{\rm E}$&\colhead{$\varepsilon_{\rm{HC}}$}&\colhead{$\varepsilon_{\rm{LC}}$}\\ & $(\rm{days})$ &$(M_{\sun})$&$(\rm{days})$&~&~&$(\rm{degree})$& & $(\rm{kpc})$& ~&~& $(R_{\star})$& $[\%]$& $[\%]$ & $[\%]$}
    \startdata
    $1$ & $27.4$ & $1.56$ & $1.87$ & $0.82$ & $0.06$ & $5.3$ & $-3.25$ & $0.13$ & $1124$ & $2.00$ & $0.52$ & $15:85$ & $4.52$&$5.55$\\
    $2$ & $54.8$ & $1.55$ & $2.40$ & $0.83$ & $0.07$ & $5.2$ & $-3.83$ & $0.13$ & $1420$ & $2.00$ & $0.53$ & $21:79$ & $5.09$&$6.06$\\
    $3$ & $82.2$ & $1.55$ & $2.87$ & $0.85$ & $0.07$ & $5.1$ & $-4.15$ & $0.13$ & $1637$ & $1.99$ & $0.53$ & $25:75$ & $5.43$&$6.29$\\
    $4$ & $109.6$ & $1.55$ & $3.11$ & $0.85$ & $0.08$ & $5.1$ & $-4.32$ & $0.13$ & $1824$ & $2.00$ & $0.53$ & $27:73$ & $5.62$&$6.46$\\
    $5$ & $137.0$ & $1.55$ & $3.43$ & $0.86$ & $0.08$ & $5.1$ & $-4.42$ & $0.13$ & $1985$ & $2.00$ & $0.54$ & $28:72$ & $5.78$&$6.56$\\
    $6$ & $164.4$ & $1.55$ & $3.74$ & $0.86$ & $0.08$ & $5.1$ & $-4.51$ & $0.13$ & $2136$ & $1.99$ & $0.54$ & $29:71$ & $5.89$&$6.66$\\
    $7$ & $191.8$ & $1.54$ & $4.20$ & $0.87$ & $0.08$ & $5.1$ & $-4.57$ & $0.14$ & $2274$ & $1.99$ & $0.54$ & $30:70$ & $5.98$&$6.75$\\
    $8$ & $219.2$ & $1.54$ & $4.45$ & $0.87$ & $0.08$ & $5.1$ & $-4.63$ & $0.14$ & $2411$ & $1.99$ & $0.54$ & $30:70$ & $6.03$&$6.79$\\
    $9$ & $246.6$ & $1.54$ & $4.64$ & $0.87$ & $0.08$ & $5.1$ & $-4.69$ & $0.14$ & $2526$ & $1.99$ & $0.54$ & $31:69$ & $6.10$&$6.85$\\
    $10$ & $274.0$ & $1.54$ & $5.06$ & $0.87$ & $0.09$ & $5.1$ & $-4.73$ & $0.14$ & $2631$ & $1.99$ & $0.54$ & $32:68$ & $6.18$&$6.86$\\
    $11$ & $301.4$ & $1.54$ & $5.13$ & $0.87$ & $0.09$ & $5.1$ & $-4.78$ & $0.14$ & $2727$ & $1.99$ & $0.54$ & $32:68$ & $6.25$&$6.89$\\
    $12$ & $328.8$ & $1.54$ & $5.36$ & $0.88$ & $0.09$ & $5.0$ & $-4.83$ & $0.14$ & $2824$ & $1.99$ & $0.54$ & $33:67$ & $6.31$&$6.92$\\
    $13$ & $356.2$ & $1.54$ & $5.36$ & $0.88$ & $0.09$ & $5.1$ & $-4.86$ & $0.14$ & $2921$ & $1.99$ & $0.54$ & $33:67$ & $6.35$&$6.97$\\
    \enddata
\end{deluxetable*}
\begin{deluxetable*}{c c c c c c c c c c c c c}
\tablecolumns{13}
\centering
\tablewidth{0.9\textwidth}\tabletypesize\footnotesize
\tablecaption{Same as Tables \ref{tab1} and \ref{tab2}, but for simulated BHMS binaries.\label{tab3}}
\tablehead{\colhead{$\rm{No}_{\rm s}$} & $T_{\rm{obs}}$&\colhead{$\overline{M_{\rm{BH}}}$}&\colhead{$\overline{T}$} & \colhead{$\overline{\log_{10}[a/R_{\star}]}$} & $\overline{\epsilon}$ & $\overline{i}$ & $\overline{D_{\rm l}}$ & $\overline{\rm{SNR}}$ & $\overline{\log_{10}[\rho_{\star}]}$& $\overline{b}$ & \colhead{$\varepsilon_{\rm{HC}}$} & \colhead{$\varepsilon_{\rm{LC}}$}\\ 
& $(\rm{days})$ &$(M_{\sun})$&$(\rm{days})$&~&~&$(\rm{degree})$ & $(\rm{kpc})$& ~& ~&$(R_{\star})$& $[\%]$& $[\%]$}
\startdata
$1$ & $27.4$ & $29.25$ & $1.24$ & $0.97$ & $0.04$ & $2.22$ & $0.13$ & $203.28$ & $1.30$ & $0.36$ & $4.15$&$4.31$\\
$2$ & $54.8$ & $29.21$ & $1.79$ & $1.00$ & $0.05$ & $2.24$ & $0.13$ & $280.08$ & $1.29$ & $0.38$ & $4.38$&$4.68$\\
$3$ & $82.2$ & $29.34$ & $2.17$ & $1.02$ & $0.05$ & $2.30$ & $0.13$ & $337.16$ & $1.28$ & $0.39$ & $4.53$&$4.82$\\
$4$ & $109.6$ & $29.45$ & $2.58$ & $1.03$ & $0.06$ & $2.33$ & $0.13$ & $382.10$ & $1.28$ & $0.40$ & $4.65$&$5.04$\\
$5$ & $137.0$ & $29.56$ & $3.06$ & $1.04$ & $0.06$ & $2.35$ & $0.13$ & $420.00$ & $1.27$ & $0.42$ & $4.76$&$5.12$\\
$6$ & $164.4$ & $29.64$ & $3.44$ & $1.05$ & $0.06$ & $2.39$ & $0.13$ & $451.62$ & $1.27$ & $0.43$ & $4.86$&$5.29$\\
$7$ & $191.8$ & $29.70$ & $3.72$ & $1.05$ & $0.06$ & $2.41$ & $0.13$ & $482.39$ & $1.26$ & $0.43$ & $4.93$&$5.40$\\
$8$ & $219.2$ & $29.72$ & $4.28$ & $1.06$ & $0.07$ & $2.44$ & $0.13$ & $508.94$ & $1.26$ & $0.45$ & $5.01$&$5.46$\\
$9$ & $246.6$ & $29.74$ & $4.52$ & $1.06$ & $0.07$ & $2.47$ & $0.13$ & $535.46$ & $1.26$ & $0.45$ & $5.07$&$5.53$\\
$10$ & $274.0$ & $29.75$ & $4.99$ & $1.07$ & $0.07$ & $2.47$ & $0.13$ & $557.64$ & $1.26$ & $0.46$ & $5.13$&$5.58$\\
$11$ & $301.4$ & $29.76$ & $5.31$ & $1.07$ & $0.07$ & $2.50$ & $0.13$ & $582.47$ & $1.25$ & $0.47$ & $5.19$&$5.63$\\
$12$ & $328.8$ & $29.81$ & $5.36$ & $1.07$ & $0.07$ & $2.53$ & $0.13$ & $602.84$ & $1.25$ & $0.47$ & $5.24$&$5.69$\\
$13$ & $356.2$ & $29.80$ & $5.56$ & $1.07$ & $0.07$ & $2.54$ & $0.13$ & $622.49$ & $1.25$ & $0.47$ & $5.29$&$5.77$\\
\enddata%\tablecomments{}
\end{deluxetable*}	
	
\begin{figure*}
\centering
\includegraphics[width=0.32\textwidth]{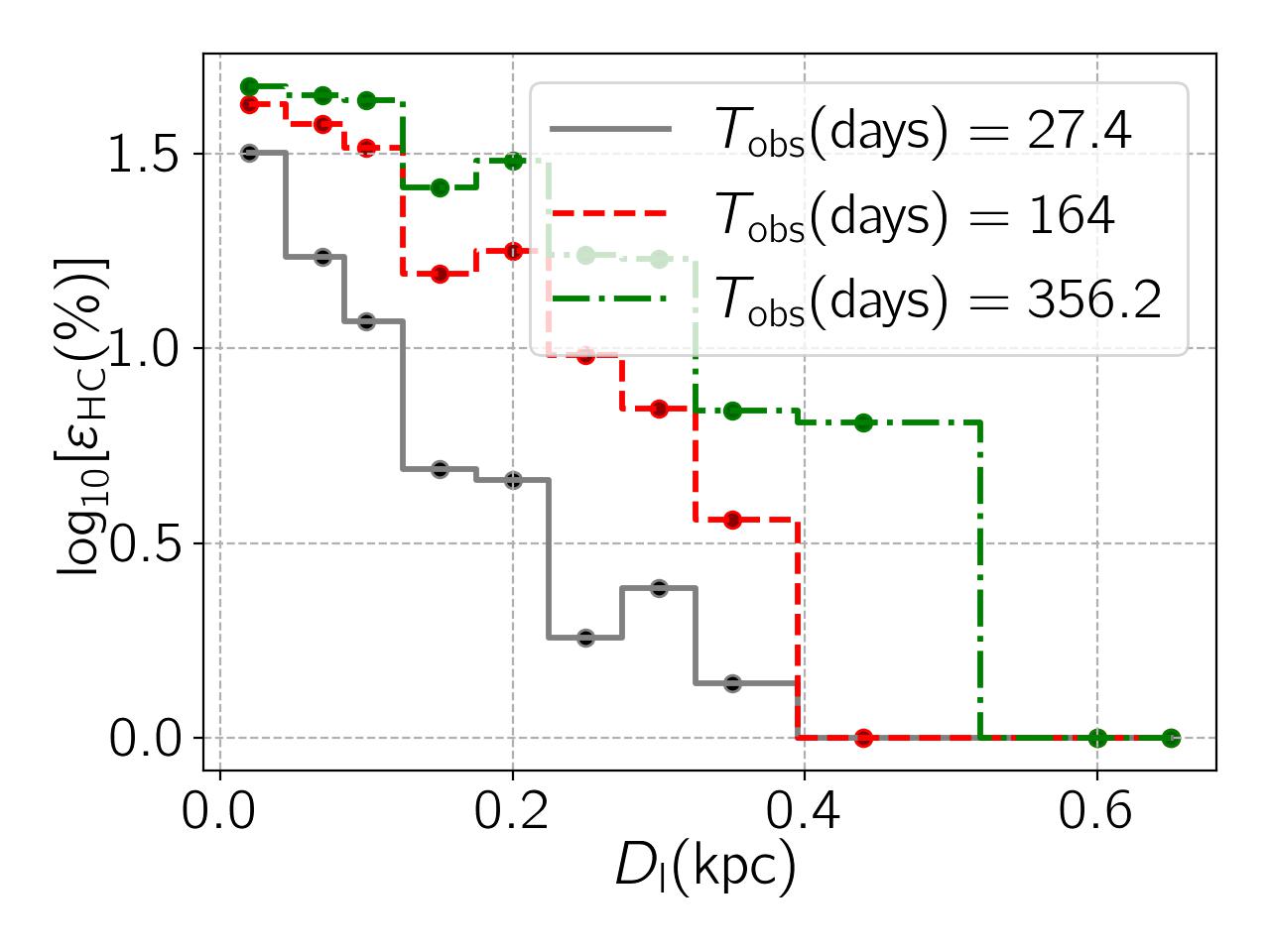}
\includegraphics[width=0.32\textwidth]{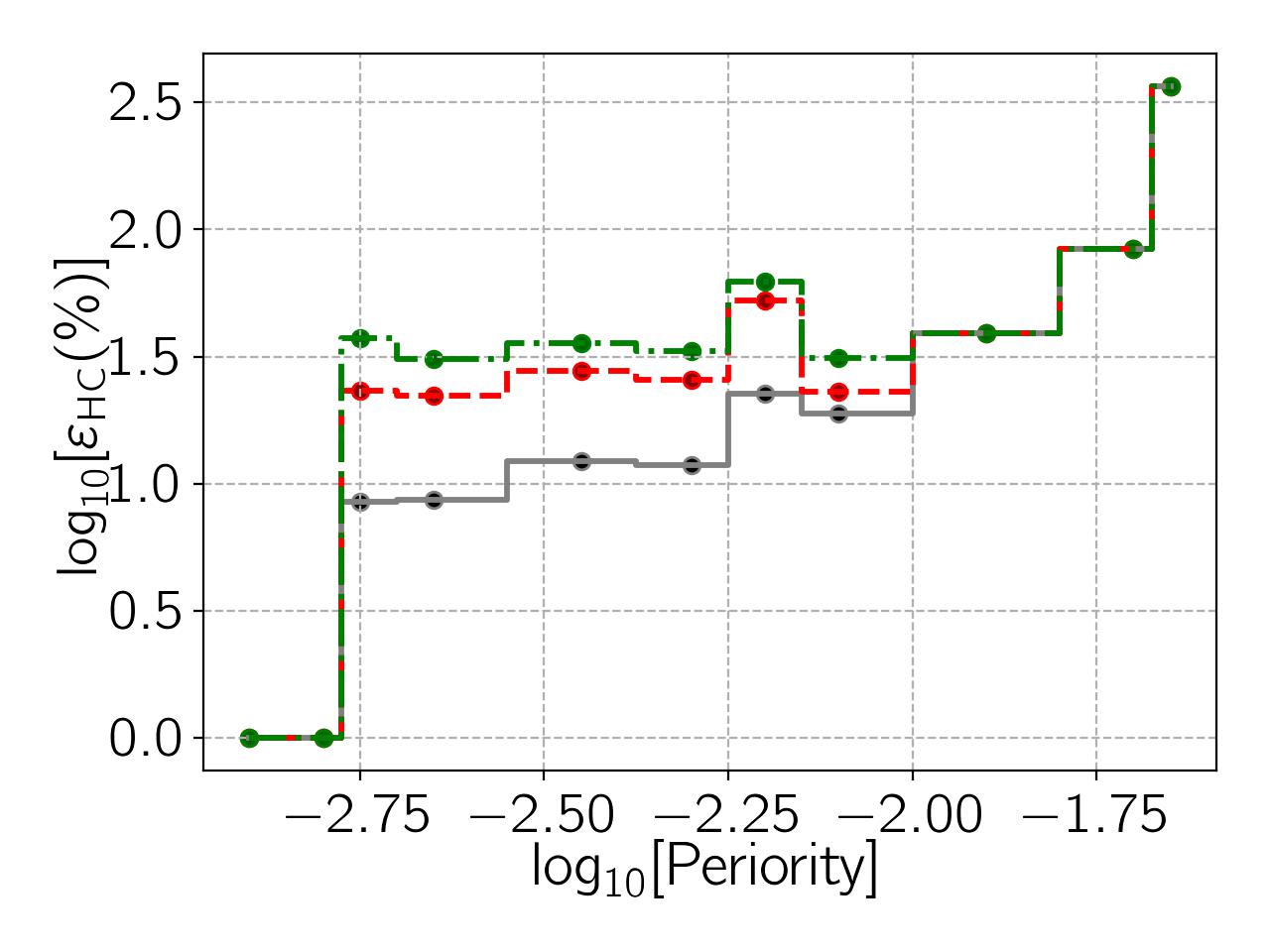}
\includegraphics[width=0.32\textwidth]{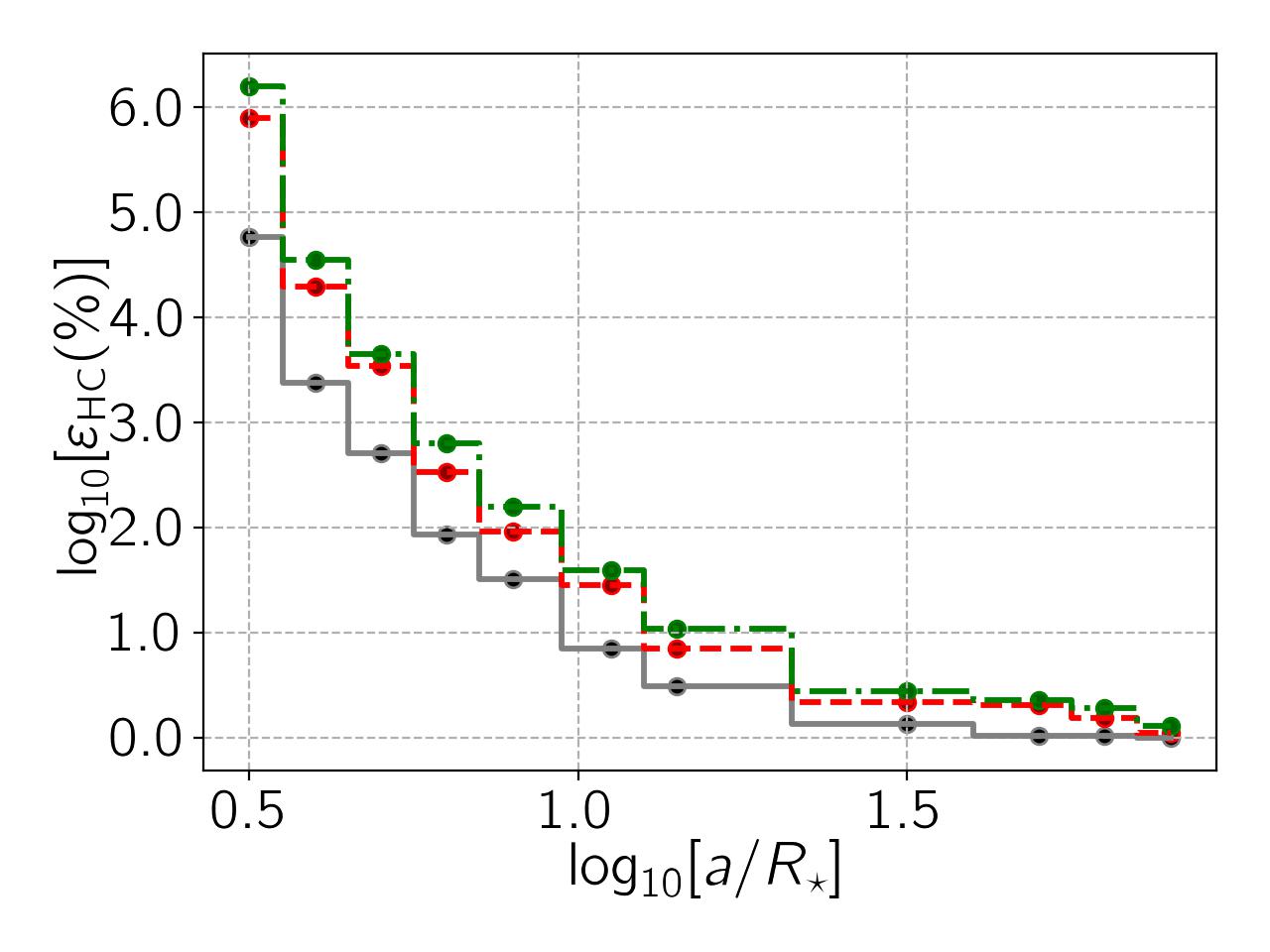}
\includegraphics[width=0.32\textwidth]{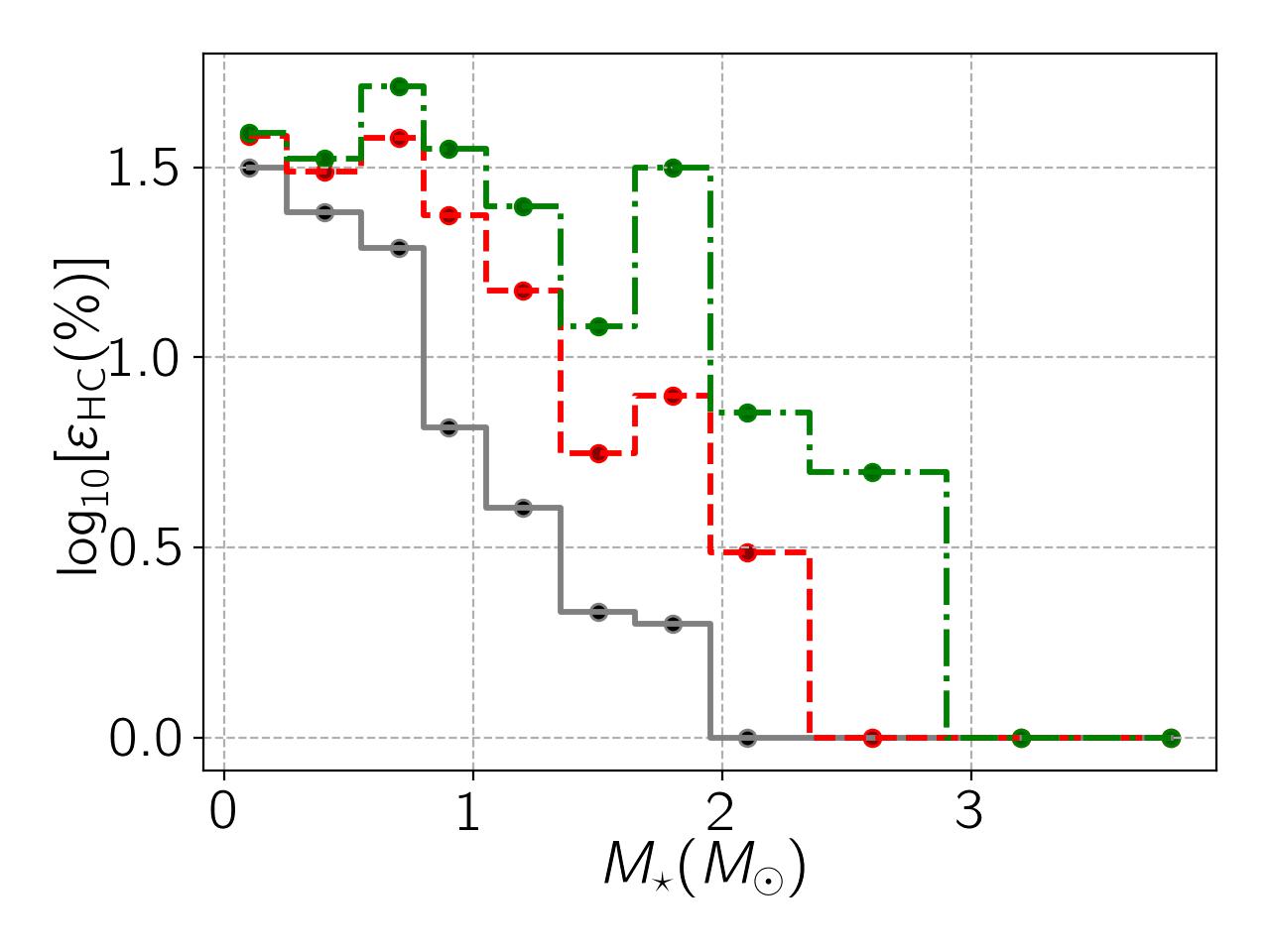}
\includegraphics[width=0.32\textwidth]{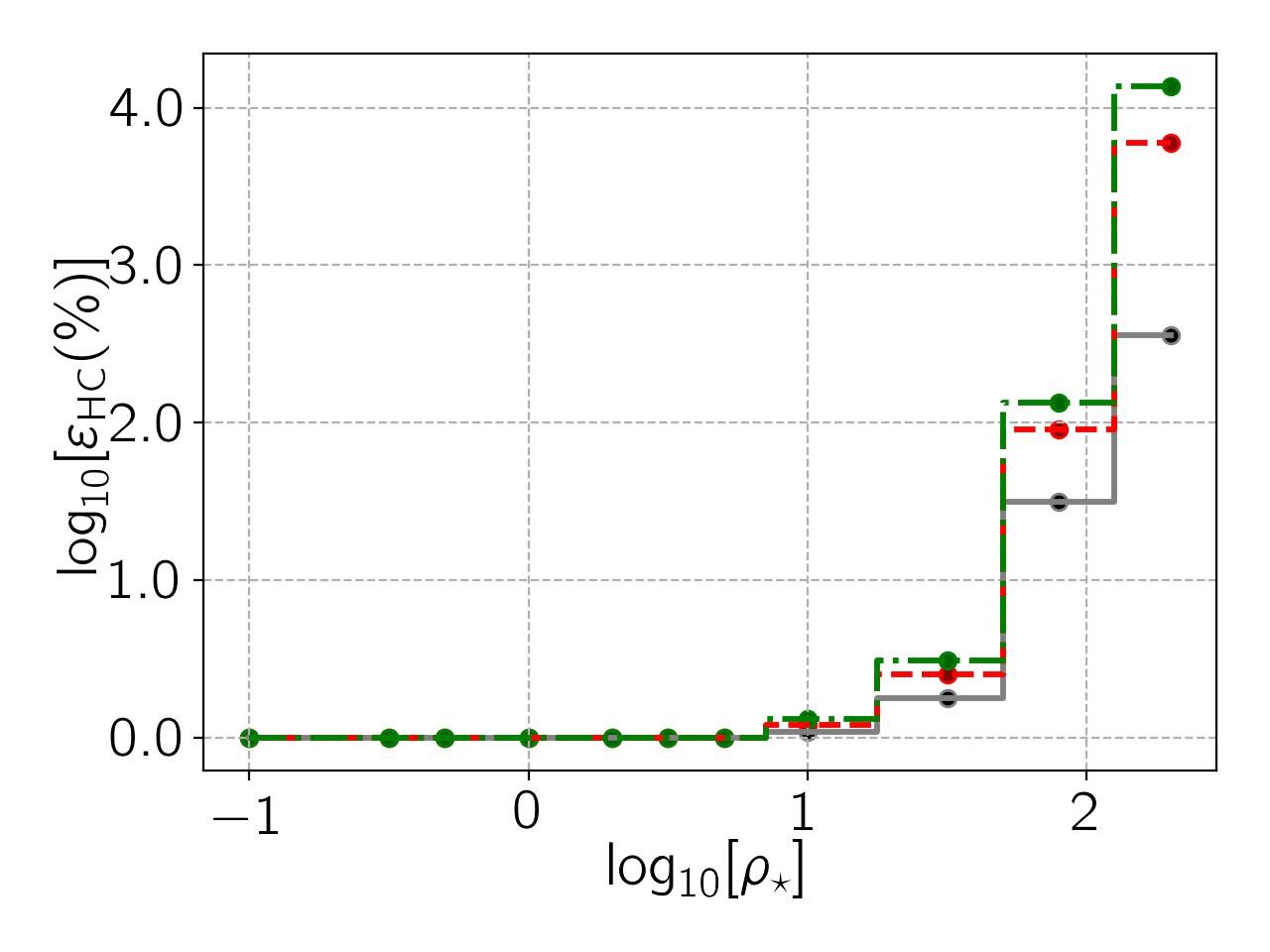}
\includegraphics[width=0.32\textwidth]{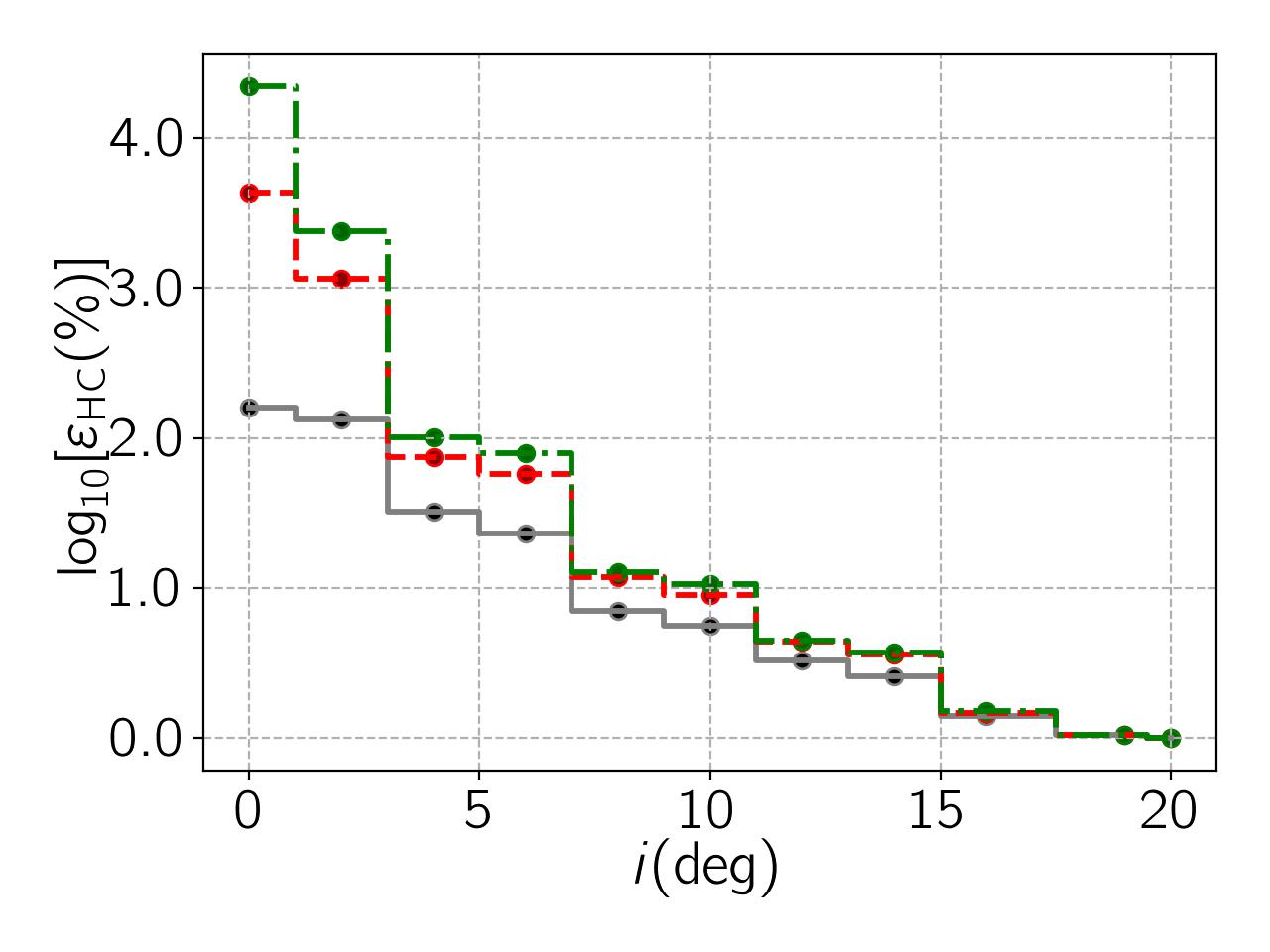}
\includegraphics[width=0.32\textwidth]{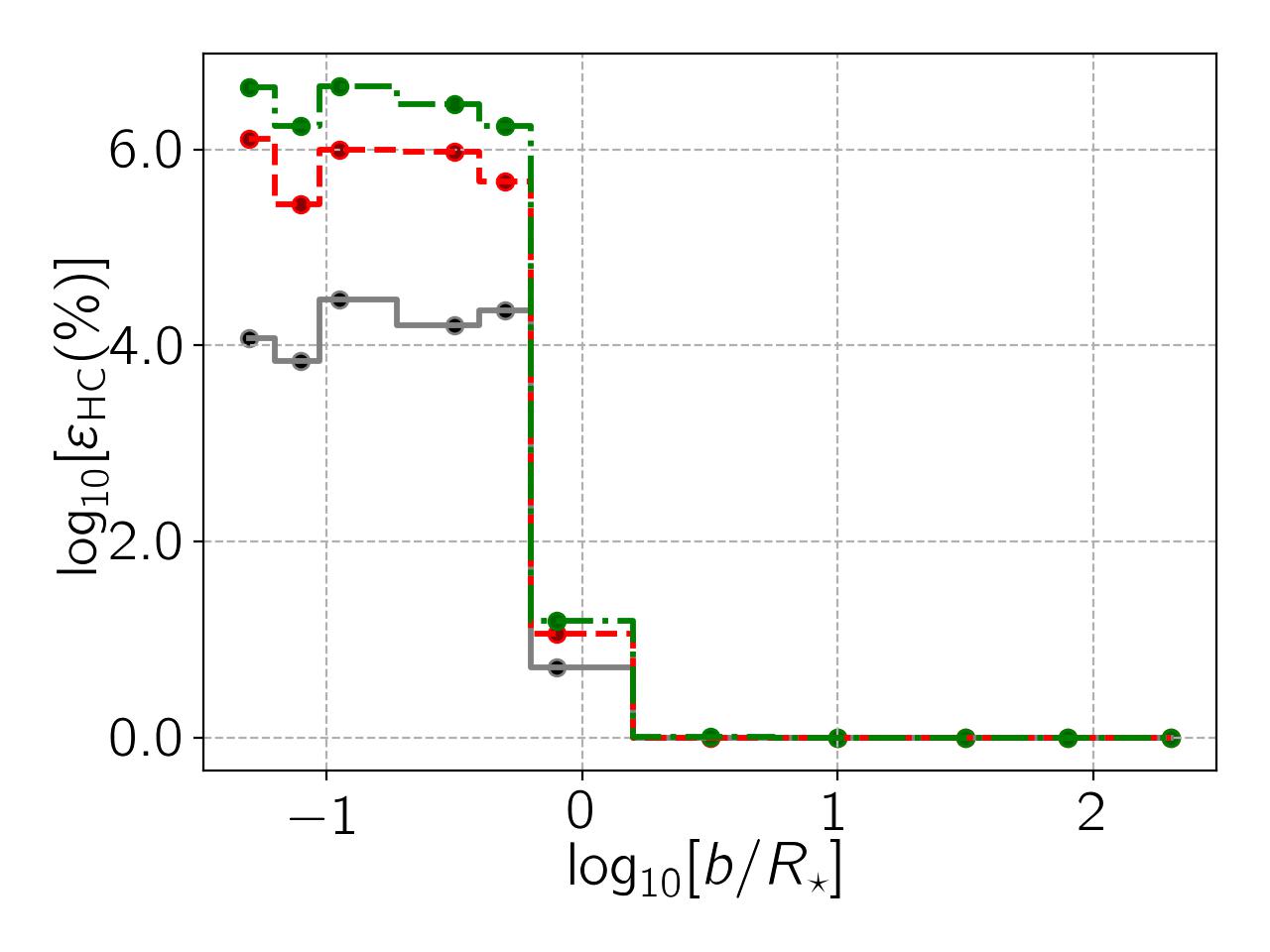}
\includegraphics[width=0.32\textwidth]{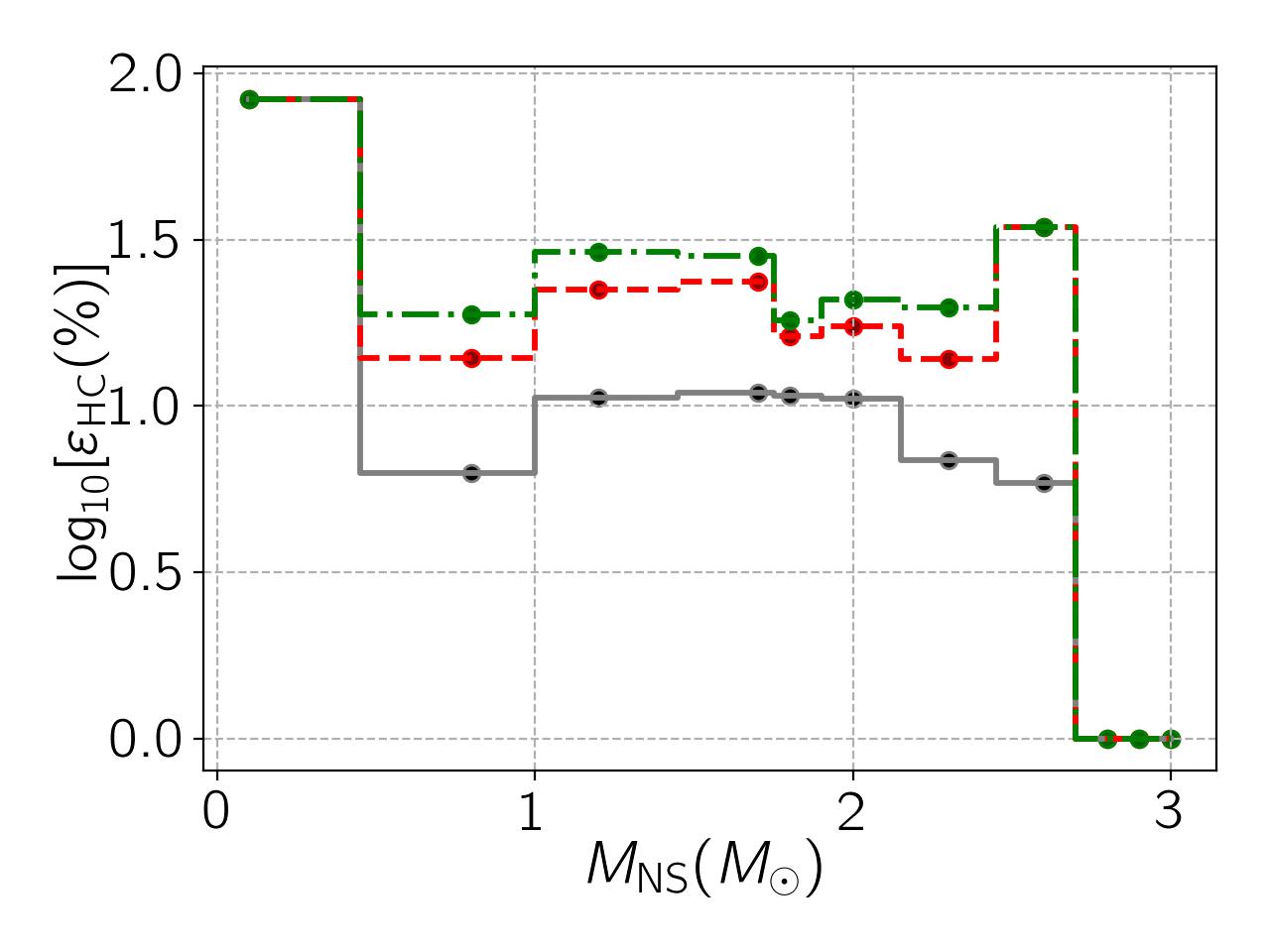}
\includegraphics[width=0.32\textwidth]{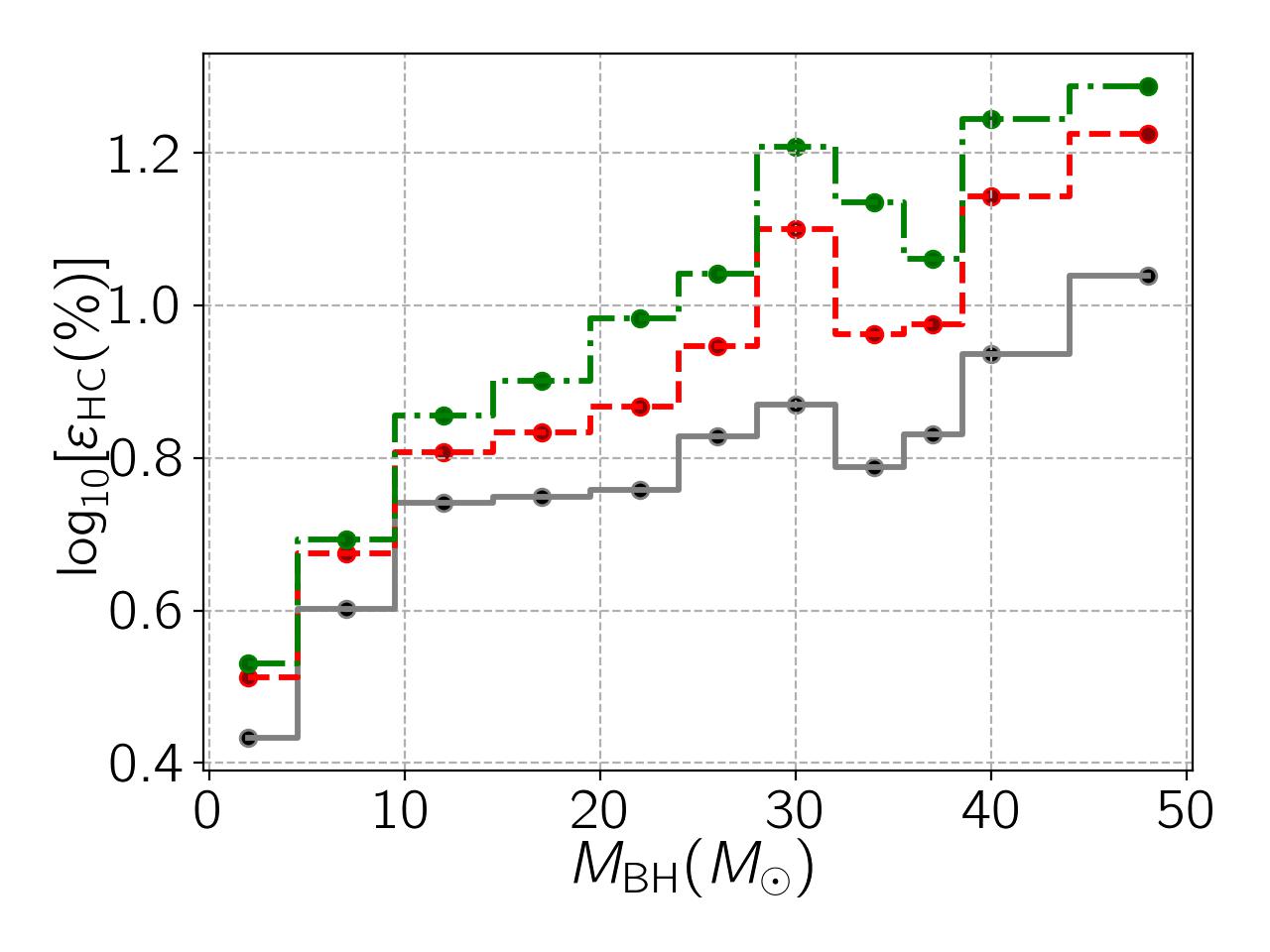}
\caption{The HC efficiencies for detecting signals due to compact objects in stellar light curves (in the logarithmic scale $\log_{10}[\varepsilon_{\rm{HC}}(\%)]$) versus nine parameters which are $D_{\rm l}(\rm{kpc})$, $\log_{10}[\rm{Periority}]$, $\log_{10}[a/R_{\star}]$, $M_{\star}(M_{\odot})$, $\log_{10}[\rho_{\star}]$, $i^{\circ}$, $\log_{10}[b/R_{\star}]$, $M_{\rm{NS}}(M_{\odot})$, and $M_{\rm{BH}}(M_{\odot})$ as shown in different panels. Three efficiency curves with different line-styles are due to three values for $T_{\rm{obs}}$ as mentioned in the first panel.}\label{Effi}
\end{figure*}

\section{Results: Statistics and Properties}\label{sec3}
We assume the maximum observational time span for a part of the sky during the TESS mission is $360$ days. Although, the southern (northern) ecliptic hemisphere was re-observed in the third (forth and fifth) year(s) of the TESS mission again, discerning periodic variations with $T(\rm{day})\geq360$ (longer than the maximum value for the continuous TESS observing time span from a part of the sky) in stellar light curves is barely possible. Because, (i) there is a $1$- up to $2$-year gap in the middle, and (ii) $74\%$ of stars are observed during only $27.4$ days of a year (they are inside one sector) and the probability of occurring either self-lensing/occultation or eclipsing signals (when the orbital period is long) exactly during that $27.4$-day observing time is low. We therefore simulate synthetic data points for the events with $T<360$ days, which means we assume all events with $T\geq 360$ days are not detectable in the TESS observations.

We perform three Monte Carlo simulations from WDMS, NSMS, and BHMS binary systems by considering different observing time spans, which can be from $27.4$ days to $356$ days due to different numbers of overlapping sectors. In Table \ref{tab1}, the results from the Monte Carlo simulation of WDMS binary systems are reported. This table includes the average values of some orbital and WDs parameters (including $M_{\rm{WD}}(M_{\sun})$, $T(\rm{days})$, $\log_{10}[a/R_{\star}]$, $\epsilon$, $i (\rm{deg})$, $\log_{10}[\mathcal{F}]$, $D_{\rm l}(\rm{kpc})$, SNR, $\log_{10}[\rho_{\star}]$, and $b(R_{\star})$) for HC detectable events during different observing time spans (given in the second column). 

\noindent Here, $b$ is the impact parameter in the lensing formalism, which is the minimum value of the projected distance $d_{\rm p}$. If this parameter is less than the source radius, eclipsing/occultation signals certainly happen. Also, this parameter determines the maximum magnification factor. The thirteenth column of this table determines (i) the fraction of simulated events which are detectable because of their self-lensing signals, $f_{\rm L}$, (ii) the fraction of simulated events which are detectable owing to their eclipsing signals $f_{\rm E}$, and (iii) the fraction of ones which are detectable due to their occultation (finite-lens effect) signals $f_{\rm O}$. Two last columns report the detection efficiencies $\varepsilon_{\rm{HC}}[\%]$ and $\varepsilon_{\rm{LC}}[\%]$ which are the ratio of detectable events (with high and low confidences, respectively) to total simulated events.

\begin{figure} 
\centering
\includegraphics[width=0.45\textwidth]{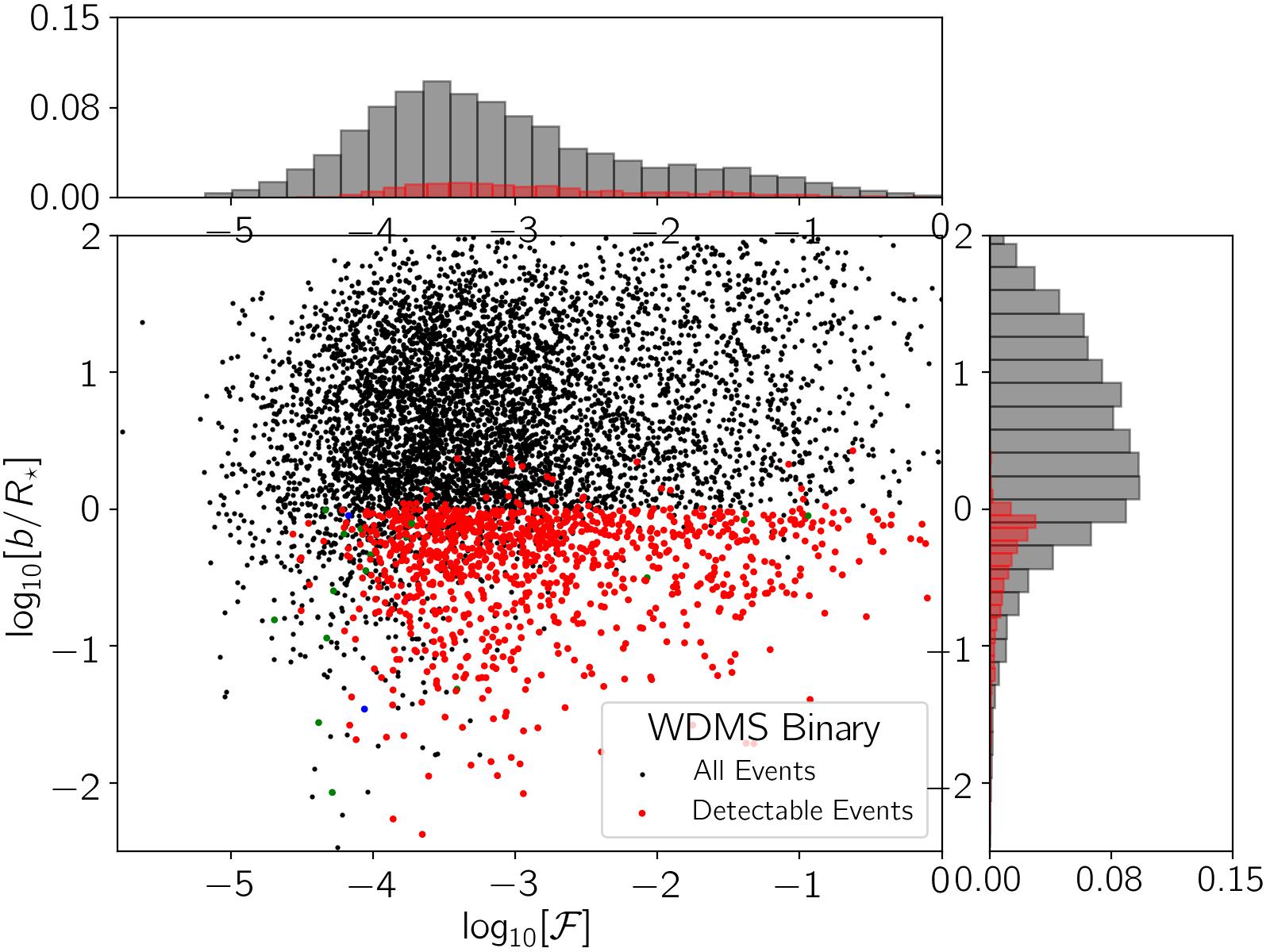}
\includegraphics[width=0.45\textwidth]{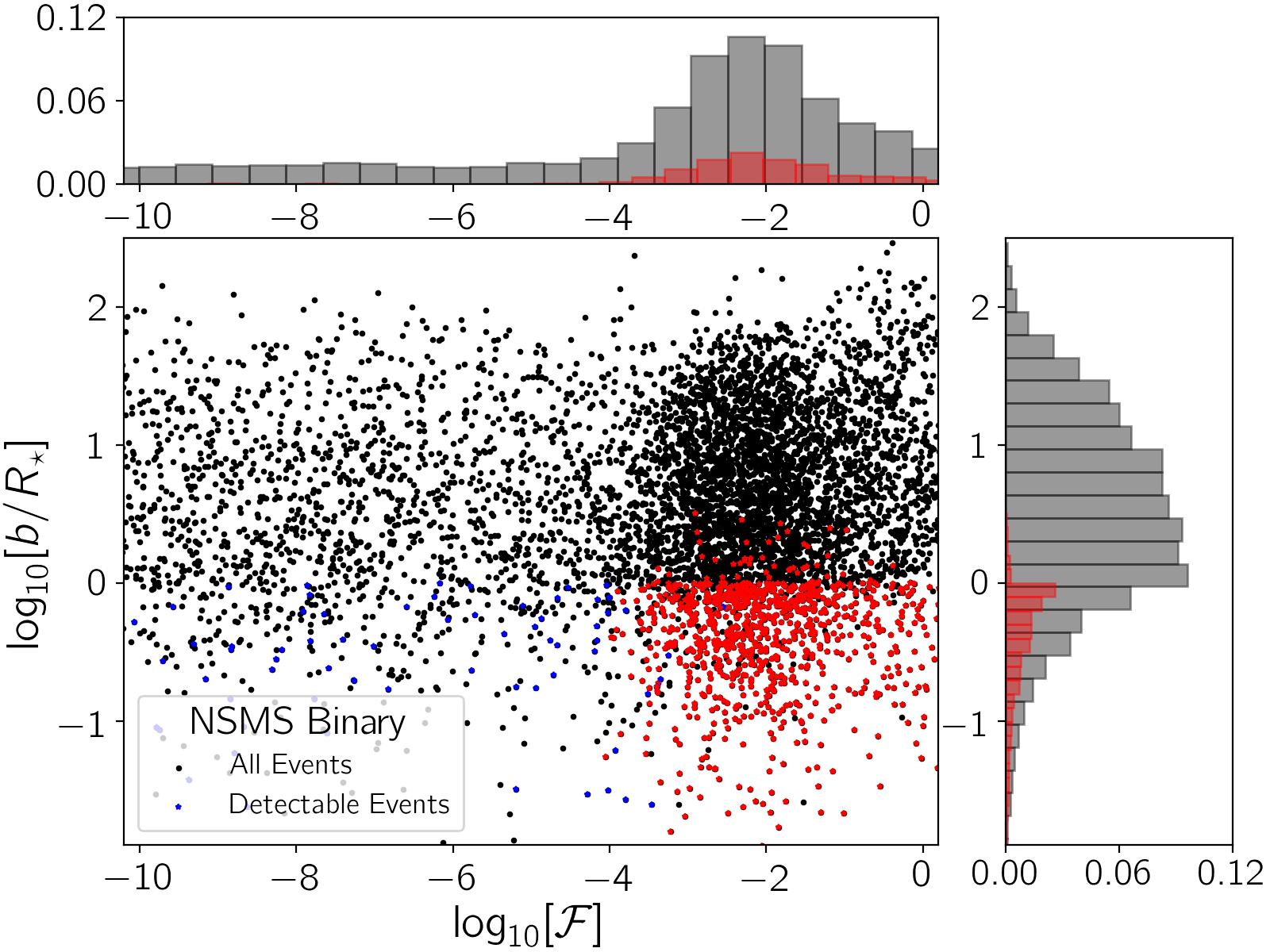}
\includegraphics[width=0.45\textwidth]{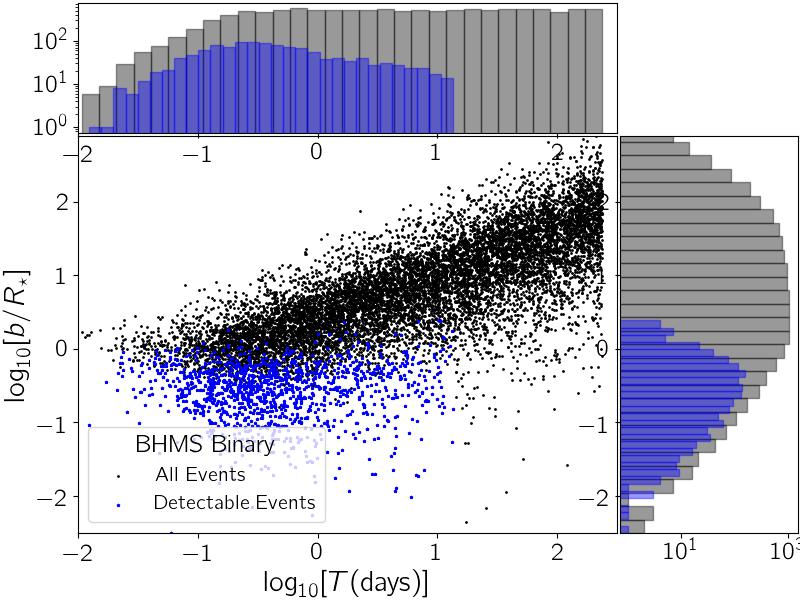}
\caption{Top panel: The scatter plot of all simulated WDMS binaries (black points) and the ones with detectable WD-induced impacts (colored points) in 2D space $\log_{10}[\mathcal{F}]-\log_{10}[b/R_{\star}]$. Their 1D and normalized distributions are shown at two sides of the plot. The red, blue, and green circles represent detectable events owing to their eclipsing, self-lensing, and occultation signals, respectively. Middle panel: Same as the previous one but resulted from the Monte Carlo simulation of NSMS binaries. Last panel: Same as two previous ones but resulted from simulating BHMS binaries and in 2D space $\log_{10}[T(\rm{days})]-\log_{10}[b/R_{\star}]$.}\label{scatter}
\end{figure}

The results from Monte Carlo simulations of detached binary systems including main-sequence stars and either NSs or SBHs are reported in Tables \ref{tab2}, and \ref{tab3}, respectively. We note that for BHMS binary systems $f_{\rm E}=0$,~$f_{\rm O}=0$, and $\mathcal{F}=0$.

Generally, longer observing time windows have two positive effects on the detectability of compact objects' signatures. For longer observing time windows, the number of transits $N_{\rm{tran}}$ is higher which (i) increases SNR values (see Eq. \ref{snr}), and (ii) enhances $N_{\rm{tran}}$ (the second detectability criterion). Enhancing $N_{\rm{tran}}$ (and accordingly SNR) for longer observing times is beneficial for detecting fainter compact objects (WDs and NSs which could be even farther), in wider and more eccentric orbits. For that reason in these three tables for longer observing times detectable compact objects are on average fainter (with less $\overline{\log_{10}[\mathcal{F}]}$) and farther. We note that due to applying the eccentricity-orbital period correlation in Monte Carlo simulations binary systems with longer orbital periods are on average more eccentric. By increasing the observing time from $27.4$ days to $356.2$ days, the detection efficiency improves by $\sim 2$.%%%-$7.5\%$.    

For $\lesssim 3\%$ and $\lesssim 33\%$ of detectable WDMS and NSMS binary systems, self-lensing signals are the most-dominant ones in stellar light curves. Indeed, for these detached WDMS and NSMS binary systems on average $\rho_{\star}\sim145,~100$ which make very flattened self-lensing signals with the depths $\Delta F_{\rm L}\sim 2 \rho^{-2}_{\star}\sim 10^{-4},~2\times 10^{-4}$, respectively, whereas their eclipsing signals, as given by $\Delta F_{\rm E}\sim \mathcal{F}\big/(1+\mathcal{F}) \sim 1$-$2\times 10^{-3},~10^{-5}$-$5\times10^{-4}$, respectively. Hence, on average for detectable WDMS and NSMS binary systems we have $\Delta F_{\rm L}\lesssim \Delta F_{\rm E}$. The occultation signal in WDMS binary systems is on average $\Delta F_{\rm O}\sim R_{\rm c}^{2}/R_{\star}^{2}\sim 10^{-4}$ which is in the same order of magnitude with the self-lensing ones. Here, we consider a common WD with the radius $R_{\rm{WD}}\sim 0.01 R_{\odot}$.

To study what kinds of binary systems and compact objects are more detectable in the TESS observations, in Figure \ref{Effi} we show the efficiency curves for detecting the impacts of compact objects with a high confidence, $\varepsilon_{\rm HC}$, in simulated stellar light curves versus nine parameters, which are $D_{\rm l}(\rm{kpc})$, $\log_{10}[\rm{Periority}]$, $\log_{10}[a/R_{\star}]$, $M_{\star}(M_{\odot})$, $\log_{10}[\rho_{\star}]$, $i(\rm{deg})$, $\log_{10}[b/R_{\star}]$, $M_{\rm{NS}}(M_{\odot})$, and $M_{\rm{BH}}(M_{\odot})$. We consider three amounts for $T_{\rm{obs}}$ as mentioned in the first panel.   

According to these plots, we conclude that closer source stars which have more priorities and higher detection efficiencies. Indeed, WDs and NSs in nearby binary systems are brighter with on average higher $\mathcal{F}$ values. Higher $\mathcal{F}$ values make deeper eclipsing signals. 

Close binary systems with smaller semi-major axes have on average shorter orbital periods, which are more suitable to be detected. Because short-period binary systems have higher $N_{\rm{tran}}$ and higher SNR values. We note that $\rho_{\star}\propto a^{-1/2}$. Therefore, although the closer binary systems have shorter orbital periods (and higher SNR values), they have larger $\rho_{\star}$s, and as a result more flattened self-lensing signals. Hence, during longer observing times wider binary systems can be detected rather via self-lensing ($f_{\rm L}$ increases with $T_{\rm{obs}}$ in Tables \ref{tab1}, and \ref{tab2}). 

The impact of stellar masses on the self-lensing signals has been shown in Figure \ref{maps}. Accordingly, less massive stars (with smaller radii) have higher self-lensing signals. Also, lower mass stars are on average fainter which results in higher $\mathcal{F}$ and deeper eclipsing signals.    

Stellar orbits with fewer inclination angles (more edge-on ones) have higher detection efficiencies because by decreasing the inclination angle the impact parameter reduces as well. The eclipsing/occultation signals can be detected only in binary systems with impact parameters less than source radii. The detection efficiency drops from $\sim 4-5\%$ to $\lesssim1\%$ when the inclination angle increases from $0$ to $15$ degrees. 

We also plot the detection efficiencies as a function of the mass of NSs and SBHs in two last panels of Figure \ref{Effi}. We note that the mass range for WDs is small, and the detection efficiency does not highly change with the mass of WDs. For NSs, we determine their luminosity according to their mass and age \citep[based on Fig. (2) of ][]{2020Potekhim}. Accordingly, by increasing the mass of NSs by $\sim 1 M_{\odot}$ the luminosity of NSs decreases up to three orders of magnitude (when they are younger than 3 million years). Hence, on average less massive NSs are brighter with higher $\mathcal{F}$ values, and deeper eclipsing signals. For detached SBHs, the only method to detect them is self-lensing. In self-lensing formalism, we have $\Delta F_{\rm L} \sim 2\rho_{\star}^{-2} \propto M_{\rm{BH}}$. Therefore, more massive SBHs make higher self-lensing signals with higher SNR values.

Top panel of Figure \ref{scatter} shows the scatter plot of simulated WDMS binaries in 2D space $\log_{10}[\mathcal{F}]-\log_{10}[b/R_{\star}]$ with black circles, with two marginal and normalized 1D distributions. The binary systems with detectable WD-induced impacts due to eclipsing, self-lensing, and occultation effects are specified with red, blue, and green circles, respectively. Accordingly, most of WDMS binaries with detectable WDs impacts have $b\lesssim R_{\star}$ and $\mathcal{F}\gtrsim 10^{-4}$. The blue points (with detectable self-lensing signals) have lower $\mathcal{F}$ up to $10^{-5}$. We therefore expect in the TESS data eclipsing-induced footprints due to WDs to be more realizable than their self-lensing/occultation impacts.  

The next panel of Figure \ref{scatter} shows a scatter plot the same as the previous one but for NSMS binary systems. Considering this point that we chose the ages of NSs uniformly from the range $\log_{10}[\mathcal{A}(\rm{year})]\in [2.5,~8]$, several NSs in our simulation are cool with very low $\mathcal{F}$ values (ones older than $\sim 1$ million years). For these systems with $\mathcal{F}\lesssim 10^{-4}$, only lensing-induced impacts are detectable (and when $b\lesssim R_{\star}$). For brighter (younger and hotter) NSs, eclipsing signals are detectable in the events with $b\lesssim R_{\star}$. We note that due to this considerable number of dim NSs in our simulation, only $f_{\rm L}\sim 15$-$33\%$ of detectable events have dominant self-lensing signals.  

The last panel of Figure \ref{scatter} shows the same plot as ones displayed in previous panels but for BHMS binary systems in 2D space $\log_{10}[T(\rm{days})]-\log_{10}[b/R_{\star}]$. The most important factor for the detectability of BH-induced lensing signals is the impact parameter $b$. Most of the simulated events with $b\lesssim R_{\star}$ have detectable self-lensing signals. We determine the impact parameter numerically from the simulation (the minimum value of $d_{\rm{p}}$), nevertheless it can be estimated as $b\simeq \tan (i) a$. Therefore, BHMS binary systems with smaller semi-major axes have smaller $b$ and shorter orbital periods which both impacts are beneficial for detecting SBHs through self-lensing signals.

Here, we estimate the number of WDs that the TESS telescope can detect through precise photometric observations from the CTL targets with a $2$-min cadence during its mission. The TESS telescope is planned to detect $1,390,486$ CTL targets during its mission\footnote{\url{https://tess.mit.edu/public/target_lists/target_lists.html}} \citep{ctlTESS}. We calculated the fractions of these CTL targets which are observed during different $T_{\rm{obs}}$ due to different $\rm{No}_{\rm s}$ values, i.e., $N_{\star}(No_{\rm s})$, as reported in the seventh column of Table (2) of \citet{sajadian2024} and extracted from Fig (2) of  \citet{2018ApJSBarsely}. These numbers are $N_{\star}(No_{\rm s})=(1031,$~$258$,~$38$,~$9$,~$5$,~$5$,~$3$,~$2$,~$1$,~$1$,~$1$,~$16$,~$13) \times 1000$. 

\noindent The number of WDMS binaries in our galaxy is predicted to be $\sim 100$ million. Considering the total number of stars in our galaxy (which is 100 billion), one thousandth ($f_{1}$) of the TESS CTL targets should be in binary systems with WDs. We note that in the simulation we limit the inclination angle to $i \in [0,~20^{\circ}]$, whereas $f_{2}=22\%$ of all binary systems should have $i\leq20^{\circ}$. We assume that the inclination angle of the orbital plane is uniformly in the range $[0,~90^{\circ}]$. By assuming that the TESS CTL targets make a common sample of stars, the number of WDs detectable in the TESS observations from the CTL targets can be estimated by:
\begin{eqnarray}
N_{\rm{WD}}\simeq f_{1} \times f_{2}\times \sum_{No_{\rm{s}}=1}^{13} N_{\star}\big(\rm{No}_{\rm{s}}\big) \times \varepsilon_{i}\big(\rm{No}_{\rm s}\big),
\end{eqnarray}
where, values of $\varepsilon_{i}\big(\rm{No}_{\rm s}\big)$ (here, $i=$ HC, and LC) are given in two last columns of Table \ref{tab1}. The estimated numbers of detectable WDs for different observing time windows and two levels of confidence are reported in two first rows of Table \ref{tab4}.

\noindent Overly, we expect $\sim 15,~18$ WDs in detached WDMS systems can be realized with high and low confidences through the TESS photometric data from the CTL targets. We note that some of these stars were (and will be) observed more than one year (two or three times) during the TESS mission. For instance, the TESS telescope observed each ecliptic hemisphere two times up to now. But, in the Monte Carlo simulations, we consider the TESS data for each CTL target taken for up to one year. Hence, these numbers are underestimations and by considering all data some more WDs in detached orbits around main-sequence stars with days will be discovered. 

\begin{deluxetable*}{c c c c c c c c c c c c c  c c}
\tablecolumns{15}
\centering
\tablewidth{0.99\textwidth}\tabletypesize\footnotesize 
\tablecaption{Estimated numbers of WDs, NSs, and SBHs with detectable impacts in the TESS CTL's light curves, by considering different observing time spans, and two levels of confidence (HC, LC).}.\label{tab4}
\tablehead{\colhead{$\rm{Number}\big/\rm{No}_{\rm s}$}&\colhead{$1$}&\colhead{$2$}&\colhead{$3$}&\colhead{$4$}&\colhead{$5$}&\colhead{$6$}&\colhead{$7$}&\colhead{$8$}&\colhead{$9$}&\colhead{$10$}&\colhead{$11$}&\colhead{$12$}&\colhead{$13$}&\colhead{$\rm{Total}$} }
\startdata
\multicolumn{15}{c}{{\bf WDs}}\\
$\rm{HC}$ &$10.52$ & $3.06$ & $0.48$ & $0.12$ & $0.07$ & $0.07$ & $0.04$ & $0.03$ & $0.01$ & $0.02$ & $0.02$ & $0.25$ & $0.20$ & $14.88$\\
$\rm{LC}$ &$12.74$ & $3.58$ & $0.56$ & $0.14$ & $0.08$ & $0.08$ & $0.05$ & $0.03$ & $0.02$ & $0.02$ & $0.02$ & $0.27$ & $0.22$ & $17.80$\\
\hline
\multicolumn{15}{c}{{\bf NSs}}\\
$\rm{HC}$ &$4.14$ & $1.17$ & $0.18$ & $0.04$ & $0.03$ & $0.03$ & $0.02$ & $0.01$ & $0.01$ & $0.01$ & $0.01$ & $0.09$ & $0.07$ & $5.80$\\
$\rm{LC}$ &$5.09$ & $1.39$ & $0.21$ & $0.05$ & $0.03$ & $0.03$ & $0.02$ & $0.01$ & $0.01$ & $0.01$ & $0.01$ & $0.10$ & $0.08$ & $7.03$\\
\hline
\multicolumn{15}{c}{{\bf SBHs}}\\
$\rm{HC}$ &$0.04$ & $0.01$ & $0.00$ & $0.00$ & $0.00$ & $0.00$ & $0.00$ & $0.00$ & $0.00$ & $0.00$ & $0.00$ & $0.00$ & $0.00$ & $0.05$\\
$\rm{LC}$ &$0.04$ & $0.01$ & $0.00$ & $0.00$ & $0.00$ & $0.00$ & $0.00$ & $0.00$ & $0.00$ & $0.00$ & $0.00$ & $0.00$ & $0.00$ & $0.05$\\
\enddata
\end{deluxetable*}

In the same way, we estimate the numbers of NSs and SBHs expected to be discovered through detecting their self-lensing and eclipsing signals in the TESS CTL's light curves which are reported in Table \ref{tab4}. The fractions of the TESS CTL targets that are in binary systems with NSs and SBHs are $f_{1}\simeq  4\times 10^{-4},~4\times 10^{-6}$, respectively. Here, we assume the binarity fractions of NSs, and SBHs (with main-sequence stars) are $4\%$ \citep{2006bookXray}. The total numbers of NSs and SBHs will be discovered with two levels of confidence through photometric observations of the TESS CTL targets are $\sim 6,~7$, and less than one, respectively.

Therefore, changing detectability criteria from a high confidence to a low one does not alter the numbers of detectable compact objects significantly. Because, according to Tables \ref{tab1}, \ref{tab2}, and \ref{tab3}, HC and LC efficiencies for detecting impacts of compact objects in WDMS, NSMS, and BHMS systems are close to each other, and $\varepsilon_{\rm{HC}}= 4.6-7.0\%,~4.5-6.4\%,~4.2-5.3\%$, and $\varepsilon_{\rm{LC}}= 5.6-7.8\%,~5.6-7.0\%,~4.3-5.8\%$. In fact, the fractions of simulated WDMS, NSMS, and BHMS binary systems with $T<360$ days are $35.3\%$, $37.3\%$, and $42.8\%$, respectively, and the corresponding fractions for systems with $T<180$ days are $32.1\%$, $33.2\%$, and $39.2\%$. Therefore, more than $\sim 60\%$ of simulated events have $T>360$ days, and the numbers of binary systems with orbital periods $\in [180,~360]$ days are not significant.

Although the number of SBHs that can be detected through self-lensing is not promising, we expect $\sim15$-$18$ WDs to be detected through their signatures in light curves of the TESS CTL targets. This number is three times larger than the number of WDs discovered from the Kepler data.  

 \section{Conclusions}
The TESS telescope observed (and observes) the CTL targets with a $2$-min cadence and a great accuracy. Although its main goal from these observations is discovering Earth-size planets transiting bright stars in the solar neighborhood, its observing strategy is uniquely matched to capture any other types of periodic and weak variations in stellar light curves. Stellar light curves from edge-on and detached binary systems including main-sequence stars and compact objects have potentially self-lensing/occultation/eclipsing signals which all are periodic. Considering three types of compact objects (i.e., WDs, NSs, and SBHs), these binary systems usually are denoted by WDMS, NSMS, and BHMS, respectively. In this work, we studied statistics and properties of WDMS, NSMS, and BHMS binary systems with detectable signatures due to compact objects in the TESS observations of the CTL targets.  

We first modeled self-lensing signals due to detached and edge-on binary systems including main-sequence stars and compact objects. The self-lensing peak, which can be estimated by $\Delta F_{\rm L}\sim 2 \rho_{\star}^{-2}$, is higher for low-mass and small stars. A self-lensing signal for a red-dwarf star with $R_{\star}=0.35 R_{\odot}$ is higher (two orders of magnitude) than that for a sub-giant star with $R_{\star}=3.6 R_{\odot}$. Peaks of self-lensing signals are degenerate functions of two parameters (i) the mass of the compact object and (ii) the source radius. Increasing the mass of compact objects and decreasing the source radius have the same effects on peaks of self-lensing signals. The inclination angle of stellar orbits around the compact objects is the only parameter which changes the shape of self-lensing signals from strict top-hat models to ones with slow-increasing edges. Self-lensing signals from eccentric stellar orbits are asymmetric concerning their peaks unless the source star is passing from either apoapsis or periapsis point while lensing.

We performed Monte-Carlo simulations from all possible stellar light curves due to detached and edge-on binary WDMS, NSMS, and BHMS systems, and assumed they are observed by the TESS telescope with a $2$-min cadence. We chose source stars in these simulations from the TESS CTL, WDs from an ensemble of $1772$ discovered nearby ones through the SDSS observations, NSs, and SBHs from their known distributions. We made synthetic data points according to the TESS observing strategy for the CTL targets and extracted the ones with the detectable signatures due to compact objects based on two sets of criteria (i) $\rm{SNR}>5$ and $N_{\rm{tran}}>2$, i.e., detecting with a high confidence(HC), and (ii) $\rm{SNR}>3$ and $N_{\rm{tran}}>1$, i.e., detecting with a low confidence (LC). %Finally, we estimated the efficiency of detecting self-lensing or eclipsing signals for these binary systems.}

\noindent There are two issues for detecting (at least one) self-lensing or eclipsing signal with the period longer than $360$ days from the TESS data which are: (i) the longest (continuous) observing time span for a part of the sky is $360$ days (happens for the ecliptic poles), and there is a gap (from one up to two years depending on the locations) between one-year missions, and (ii) around $74\%$ of the CTL targets are observed during only $27.4$ days of a one-year mission. We therefore assumed that the TESS detection efficiency for $T\geq360$ days is zero, although it is actually close to zero.

We found that the probability that simulated binary WDMB, NSMS, and BHMS systems have $T<180(360)$ days were $32.1(35.3)\%$, $33.2(37.3)\%$, and $39.2(42.8)\%$, respectively. The HC (and LC) efficiencies for detecting periodic signatures from WDMS, NSMS, and BHMS binaries during different observing time spans of a one-year mission of TESS, (i.e., $T_{\rm{obs}}=27.4 \times No_{\rm{s}}$, where $No_{\rm s}=1,~2,~3,...,~13$) are $4.6-7.0 (5.6-7.8)\%$, $4.5-6.4 (5.6-7.0)\%$, and $4.2-5.3 (4.3-5.8)\%$, respectively. Increasing observing time spans improves the detection efficiency by $\sim 2\%$ and fainter compact objects in wider and more eccentric orbits can be detected during longer observing time spans (see Tables \ref{tab1}, \ref{tab2}, and \ref{tab3}). The fractions of detectable WDMS and NSMS binary events which their self-lensing signals are the most-dominant ones are $\lesssim 3\%$, and $\lesssim 33\%$, respectively.  

We found that the detection efficiency is higher for closer CTL targets with higher priorities and smaller radii. Most of binary systems with detectable periodic signals have the impact parameters $b \lesssim R_{\star}$. The detection efficiencies for detecting more massive NSs, and SBHs is higher.  
 
We estimated the total number of WDs, NSs, and SBHs that can be discovered from the TESS CTL observations which are $15-18$, $6-7$, and less than one, respectively. The number of detectable WDs in the TESS data is three times higher than the number of WDs discovered in the Kepler data. \\

\small
{All simulations that have been done for this paper are available at:  \url{https://github.com/SSajadian54/SelfLensing_Eclipsing_simulator}. The codes, and several examples of generated light curves can be found in the Zenodo repository\citep{2024_zenodo}.}\\

\small{In Monte-Carlo simulations, we use the TESS CTL (with DOI number: \url{doi:10.17909/fwdt-2x66}) that are publicly available from the MAST catalog. Funding for the TESS mission is provided by NASA's Science Mission directorate. We acknowledge the use of TESS Alert data, which is currently in a beta test phase, from pipelines at the TESS Science Office and at the TESS Science Processing Operations Center. %The author gratefully thanks the anonymous referee for his/her careful and useful comments. 
NA is supported by the University of Waterloo, Natural Sciences and Engineering Research Council of Canada (NSERC) and the Perimeter Institute for Theoretical Physics. Research at Perimeter Institute is supported in part by the Government of Canada through the Department of Innovation, Science and Economic Development Canada and by the Province of Ontario through the Ministry of Colleges and Universities.}\\

\bibliographystyle{aasjournal}
\bibliography{ref}{}
\end{document}